%% file: mainALICE.tex
\documentclass[ALICE,manyauthors]{cernphprep}
\usepackage[comma,square,numbers,sort&compress]{natbib}
\usepackage{hyperref}
\usepackage{lineno}
\usepackage{xspace}
\usepackage{xcolor}
\usepackage{amssymb}
\usepackage{booktabs}
\usepackage{csquotes}
\usepackage{comment}
\usepackage{subfig}
\usepackage[T1]{fontenc}

\begin{document}
\input{commands.tex}

\begin{titlepage}
\PHyear{2021}       
\PHnumber{103}      
\PHdate{7 June}  

\title{Measurement of prompt $\Dzero$, $\Lambdac$, and $\Sigmac^{0,++}(2455)$ production\\ in proton--proton collisions at $\s=13$ TeV}
\ShortTitle{\small Measurement of prompt $\Dzero$, $\Lambdac$, and $\Sigmac^{0,++}$ production in pp collisions at $\s=13$ TeV}   

\Collaboration{ALICE Collaboration\thanks{See Appendix~\ref{app:collab} for the list of collaboration members}}
\ShortAuthor{ALICE Collaboration} 

\begin{abstract}
\input{Abstract}
\end{abstract}
\end{titlepage}

\setcounter{page}{2} 

\input{AllTogether}


\newenvironment{acknowledgement}{\relax}{\relax}
\begin{acknowledgement}
\section*{Acknowledgements}
\input{fa_2021-05-12.tex}
\end{acknowledgement}

\bibliographystyle{utphys}   
\bibliography{bibliography}

\newpage
\appendix

%
%

\section{The ALICE Collaboration}
\label{app:collab}
\input{2021-05-12-Alice_Authorlist_2021-05-12.tex}  
\newpage
\input{SupplementalMat}
\end{document}

%% file: commands.tex
%

\newcommand{\pp}           {pp\xspace}
\newcommand{\ppbar}        {\mbox{$\mathrm {p\overline{p}}$}\xspace}
\newcommand{\XeXe}         {\mbox{Xe--Xe}\xspace}
\newcommand{\PbPb}         {\mbox{Pb--Pb}\xspace}
\newcommand{\pA}           {\mbox{pA}\xspace}
\newcommand{\pPb}          {\mbox{p--Pb}\xspace}
\newcommand{\AuAu}         {\mbox{Au--Au}\xspace}
\newcommand{\dAu}          {\mbox{d--Au}\xspace}

\newcommand{\s}            {\ensuremath{\sqrt{s}}\xspace}
\newcommand{\snn}          {\ensuremath{\sqrt{s_{\mathrm{NN}}}}\xspace}
\newcommand{\pt}           {\ensuremath{p_{\rm T}}\xspace}
\newcommand{\meanpt}       {$\langle p_{\mathrm{T}}\rangle$\xspace}
\newcommand{\ycms}         {\ensuremath{y_{\rm CMS}}\xspace}
\newcommand{\ylab}         {\ensuremath{y_{\rm lab}}\xspace}
\newcommand{\etarange}[1]  {\mbox{$\left | \eta \right |~<~#1$}}
\newcommand{\yrange}[1]    {\mbox{$\left | y \right |~<~#1$}}
\newcommand{\dndy}         {\ensuremath{\mathrm{d}N_\mathrm{ch}/\mathrm{d}y}\xspace}
\newcommand{\dndeta}       {\ensuremath{\mathrm{d}N_\mathrm{ch}/\mathrm{d}\eta}\xspace}
\newcommand{\avdndeta}     {\ensuremath{\langle\dndeta\rangle}\xspace}
\newcommand{\dNdy}         {\ensuremath{\mathrm{d}N_\mathrm{ch}/\mathrm{d}y}\xspace}
\newcommand{\Npart}        {\ensuremath{N_\mathrm{part}}\xspace}
\newcommand{\Ncoll}        {\ensuremath{N_\mathrm{coll}}\xspace}
\newcommand{\dEdx}         {\ensuremath{\textrm{d}E/\textrm{d}x}\xspace}
\newcommand{\RpPb}         {\ensuremath{R_{\rm pPb}}\xspace}

\newcommand{\nineH}        {$\sqrt{s}~=~0.9$~Te\kern-.1emV\xspace}
\newcommand{\seven}        {$\sqrt{s}~=~7$~Te\kern-.1emV\xspace}
\newcommand{\twoH}         {$\sqrt{s}~=~0.2$~Te\kern-.1emV\xspace}
\newcommand{\twosevensix}  {$\sqrt{s}~=~2.76$~Te\kern-.1emV\xspace}
\newcommand{\five}         {$\sqrt{s}~=~5.02$~Te\kern-.1emV\xspace}
\newcommand{\twosevensixnn}{$\sqrt{s_{\mathrm{NN}}}~=~2.76$~Te\kern-.1emV\xspace}
\newcommand{\fivenn}       {$\sqrt{s_{\mathrm{NN}}}~=~5.02$~Te\kern-.1emV\xspace}
\newcommand{\LT}           {L{\'e}vy-Tsallis\xspace}
\newcommand{\GeVc}         {Ge\kern-.1emV/$c$\xspace}
\newcommand{\MeVc}         {Me\kern-.1emV/$c$\xspace}
\newcommand{\TeV}          {Te\kern-.1emV\xspace}
\newcommand{\GeV}          {Ge\kern-.1emV\xspace}
\newcommand{\MeV}          {Me\kern-.1emV\xspace}
\newcommand{\GeVmass}      {Ge\kern-.2emV/$c^2$\xspace}
\newcommand{\MeVmass}      {Me\kern-.2emV/$c^2$\xspace}
\newcommand{\lumi}         {\ensuremath{\mathcal{L}}\xspace}

\newcommand{\ITS}          {\rm{ITS}\xspace}
\newcommand{\TOF}          {\rm{TOF}\xspace}
\newcommand{\ZDC}          {\rm{ZDC}\xspace}
\newcommand{\ZDCs}         {\rm{ZDCs}\xspace}
\newcommand{\ZNA}          {\rm{ZNA}\xspace}
\newcommand{\ZNC}          {\rm{ZNC}\xspace}
\newcommand{\SPD}          {\rm{SPD}\xspace}
\newcommand{\SDD}          {\rm{SDD}\xspace}
\newcommand{\SSD}          {\rm{SSD}\xspace}
\newcommand{\TPC}          {\rm{TPC}\xspace}
\newcommand{\TRD}          {\rm{TRD}\xspace}
\newcommand{\VZERO}        {\rm{V0}\xspace}
\newcommand{\VZEROA}       {\rm{V0A}\xspace}
\newcommand{\VZEROC}       {\rm{V0C}\xspace}
\newcommand{\Vdecay} 	   {\ensuremath{V^{0}}\xspace}

\newcommand{\ee}           {\ensuremath{e^{+}e^{-}}} 
\newcommand{\pip}          {\ensuremath{\pi^{+}}\xspace}
\newcommand{\pim}          {\ensuremath{\pi^{-}}\xspace}
\newcommand{\kap}          {\ensuremath{\rm{K}^{+}}\xspace}
\newcommand{\kam}          {\ensuremath{\rm{K}^{-}}\xspace}
\newcommand{\pbar}         {\ensuremath{\rm\overline{p}}\xspace}
\newcommand{\kzero}        {\ensuremath{{\rm K}^{0}_{\rm{S}}}\xspace}
\newcommand{\lmb}          {\ensuremath{\Lambda}\xspace}
\newcommand{\almb}         {\ensuremath{\overline{\Lambda}}\xspace}
\newcommand{\Om}           {\ensuremath{\Omega^-}\xspace}
\newcommand{\Mo}           {\ensuremath{\overline{\Omega}^+}\xspace}
\newcommand{\X}            {\ensuremath{\Xi^-}\xspace}
\newcommand{\Ix}           {\ensuremath{\overline{\Xi}^+}\xspace}
\newcommand{\Xis}          {\ensuremath{\Xi^{\pm}}\xspace}
\newcommand{\Oms}          {\ensuremath{\Omega^{\pm}}\xspace}
\newcommand{\degree}       {\ensuremath{^{\rm o}}\xspace}

\newcommand{\XicZero}{\rm \Xi_{c}^{0}}
\newcommand{\XicPlus}{\rm \Xi_{c}^{+}}
\newcommand{\eepm}{\rm e^{+}e^{-}}
\newcommand{\Dzero}{\rm D^{0}}
\newcommand{\Dplus}{\rm D^+}
\newcommand{\Dstar}{\rm D^{*+}}
\newcommand{\Ds}{\rm D_s}
\newcommand{\DtoKpi}{\rm D^0\to K^-\pi^+}
\newcommand{\Sigmac}{\rm \Sigma_{c}}
\newcommand{\SigmacZeroPlusPlus}{{\rm \Sigma_{c}^{0,++}}}
\newcommand{\SigmacZero}{{\rm \Sigma_{c}^{0}}}
\newcommand{\SigmacPlusPlus}{{\rm \Sigma_{c}^{++}}}
\newcommand{\SigmacZeroPlusPlusPlus}{\rm \Sigma_{c}^{0,+,++}}
\newcommand{\Lambdac}{\rm \Lambda_{c}^{+}}
\newcommand{\LambdacTopKpi}{\rm \Lambda_{c}^+\rightarrow pK^{-}\pi^{+}}
\newcommand{\LambdacTopKzeroS}{\rm \Lambda_{c}^{+}\rightarrow pK^{0}_{S}}
\newcommand{\LambdacFromScZeroPlusPlusPlus}{\Lambda_{\rm c}^+\leftarrow~\Sigma_{\rm c}^{0,+,++}}
\newcommand{\LambdacFromSc}{\Lambda_{\rm c}\leftarrow\Sigma_{\rm c}}
\newcommand{\LambdacFromScZeroPlusPlus}{\Lambda_{\rm c}^{+}\leftarrow\Sigma_{\rm c}^{0,++}}
\newcommand{\ptLcSc}{p_{\text{T}}^{\LambdacFromSc}}
\newcommand{\ptSc}{p_{\text{T}}^{\Sigmac}}
\newcommand{\ptScSigned}{p_{\text{T}}^{\Sigma_{\text{c}}^{0,++}}}
\newcommand{\ptLc}{p_{\text{T}}^{\Lambdac}}
\newcommand{\Lambdab}{\rm \Lambda_{b}^{0}}
\newcommand{\ep}{\rm ep}
\newcommand{\pKpi}{\rm pK^{-}\pi^{+}}
\newcommand{\pKzeros}{\rm pK^{0}_{s}}
\newcommand{\gevc}         {{\rm GeV}/c}
\newcommand{\mevc}         {{\rm MeV}/c}
\newcommand{\DeltaM}{\Delta M}
\newcommand{\tev}          {\rm TeV}
\newcommand{\gev}          {\rm GeV}
\newcommand{\mev}          {\rm MeV}
\newcommand{\sqrts}{\sqrt{s}}
\newcommand{\SigmacTopKpipi}{\rm \SigmacZeroPlusPlus\rightarrow pK^{-}\pi^{+}\pi^{-,+}}
\newcommand{\SigmacTopKzeropiS}{\rm \SigmacZeroPlusPlus\rightarrow pK^{0}_{S}\pi^{-,+}}
\newcommand{\noop}[1]{}

%% file: Abstract.tex
The $\pt$-differential production cross sections of prompt $\Dzero$, $\Lambdac$, and $\SigmacZeroPlusPlus(2455)$ charmed hadrons are measured at midrapidity ($|y|<0.5$) in pp collisions at $\sqrt{s}=13$~TeV. This is the first measurement of $\SigmacZeroPlusPlus$ production in hadronic collisions. Assuming the same production yield for the three $\SigmacZeroPlusPlusPlus$ isospin states, the baryon-to-meson cross section ratios $\SigmacZeroPlusPlusPlus/\Dzero$ and $\Lambdac/\Dzero$ are calculated in the transverse momentum ($p_{\rm T}$) intervals $2<p_{\rm T}<12$~GeV/{\it c} and $1<p_{\rm T}<24$ ~GeV/{\it c}. Values significantly larger than in $\eepm$ collisions are observed, indicating for the first time that baryon enhancement in hadronic collisions also extends to the $\Sigmac$. The feed-down contribution to $\Lambdac$ production from $\SigmacZeroPlusPlusPlus$ is also reported and is found to be larger than in $\eepm$ collisions. The data are compared with predictions from event generators and other phenomenological models, providing a sensitive test of the different charm-hadronisation mechanisms implemented in the models.

%% file: AllTogether.tex
\newpage
The formation of hadrons out of quarks (“hadronisation”) represents a fundamental process in nature that can be investigated at particle colliders where, at high collision energies, quarks represent the relevant degrees of freedom for a very short time of the order of $10^{-23}$~s. The measurement of the relative production rates of different charm hadron species allows to study how charm quarks, produced only in initial hard scatterings, combine with other quarks, which may either exist in the system before hadronisation or be produced in the strong-force potential at hadronisation time. Recent measurements of $\Lambdac$-, $\XicZero$- and $\Lambdab$-baryon production in pp collisions at $\sqrts~=~5.02$, 7, and 13\,\TeV~\cite{Acharya:2017kfy,Acharya:2017lwf,Acharya:2020uqi,Acharya:2020lrg,Sirunyan:2019fnc,Aaij:2013mga,Aaij:2015fea,Aaij:2019pqz} indicate that the production of charm and beauty baryons relative to that of charm and beauty mesons is enhanced in pp with respect to $\eepm$ and $\ep$ collisions~\cite{Barate:1999bg,Alexander:1996wy,Abreu:1999vw,Gladilin:2014tba,Chekanov:2005mm,Abramowicz:2010aa,Abramowicz:2013eja}. %
Several models tuned to reproduce the $\eepm$ data significantly underestimate the ratios measured in pp collisions and do not describe the observed transverse-momentum ($\pt$) trends. %
These measurements also set kinematic boundaries to the validity of the assumption made in perturbative-QCD calculations like FONLL~\cite{Cacciari:1998it,Cacciari:2012ny} and GM-VFNS~\cite{Kniehl:2004fy,Kniehl:2012ti,Benzke:2017yjn,Kramer:2017gct,Helenius:2018uul,Kniehl:2020} that fragmentation functions tuned on $\eepm$ and $\ep$ data can be used in \pp collisions.

The $\SigmacZeroPlusPlusPlus$ baryon triplet is the isospin $I=1$ partner of the singlet ($I=0$) $\Lambdac$ baryon. All these states are composed of a charm quark and a pair of light (u, d) quarks. In $\eepm$ collisions, while in the light-flavour sector the mass dependence of the yields of the $\Sigma$ and $\Lambda$ states is well described by a single exponential function, the yields of the $\SigmacZeroPlusPlusPlus$ states are about a factor 4 smaller than those of the $\Lambdac$-states~\cite{Niiyama:2017wpp}. 
In the framework of hadronisation via string fragmentation, this suppression can be ascribed to the need to form $\SigmacZeroPlusPlusPlus$ via the combination of a heavy charm quark, which is always a string endpoint, and a diquark with spin $S=1$ and $I=1$ formed via the Schwinger tunnelling process~\cite{Christiansen:2015yqa,Niiyama:2017wpp}. The large mass of $S=1$ diquarks suppresses their formation with respect to $S=0$ diquarks, hence the $\SigmacZeroPlusPlusPlus$ production yield is suppressed with respect to the $\Lambdac$ yield. In the models that provide a fair description of the $\Lambdac/\Dzero$ ratio in pp collisions~(here denoted as \enquote{CR-BLC}~\cite{Christiansen:2015yqa}, \enquote{SHM+RQM}~\cite{He:2019tik}, \enquote{Catania}~\cite{Plumari:2017ntm,Minissale:2020bif}, \enquote{QCM}~\cite{Song:2018tpv}) this suppression mechanism is absent or heavily reduced, and a sizeable contribution to $\Lambdac$ production from strong decays of $\SigmacZeroPlusPlusPlus$ states is expected. Therefore, the measurement of the ground-state $\SigmacZeroPlusPlusPlus(2455)$ production is fundamental to understand the dynamics of heavy-flavour baryon formation, providing a key test for the different scenarios proposed in the mentioned models. Among these, the CR-BLC model is a version of PYTHIA 8 in which terms beyond the leading-colour approximation (BLC) are considered in string formation, representing more accurately the QCD SU(3) algebra and de facto enhancing effects from colour reconnection (CR). These terms cause confining potentials to also arise between quarks not produced in the same hard scattering and are relevant to hadronic collisions at high energies, where multiple-parton interactions produce an environment rich in quarks and gluons. Moreover, they give rise to \enquote{junction topologies} that favour the production of baryon states and do not penalise the formation of  $\SigmacZeroPlusPlusPlus$ with respect to $\Lambdac$ states. 
The production of $\SigmacZeroPlusPlusPlus$(2455) is expected to increase by large factors, up to 25, and become even larger than that of direct $\Lambdac$. The SHM+RQM model 
predicts a large feed-down contribution to the $\Lambdac$ ground state from an enriched set of mostly unobserved excited charm-hadron states expected from the Relativistic Quark Model (RQM~\cite{Ebert:2011kk}). The branching fractions of charm quarks to the various hadron species are assumed to follow the relative thermal densities calculated with the Statistical Hadronisation Model (SHM~\cite{Andronic:2003zv}), therefore to depend only on the state mass and spin-degeneracy factor. In the Catania model charm quarks can hadronise via \enquote{vacuum}-like fragmentation as well as recombine (coalesce) with surrounding light quarks from the underlying event. The Wigner formalism is used to calculate the probability to form a baryon (meson) given the phase-space distribution of three (two) quarks. A different formalism is implemented in the QCM (\enquote{quark (re-)combination mechanism}) model, in which charm quarks form hadrons by combining with equal-velocity light quarks. In this model, the relative abundances of the different baryon species are fixed by thermal weights.

In this letter, the measurement performed with the ALICE experiment of the $\pt$-differential cross sections of prompt $\Dzero$, $\Lambdac$, and $\Sigmac^{0,++}(2455)$ in pp collisions at $\sqrt{s}=13$\,\TeV at midrapidity ($|y|<0.5$) is reported. This is the first production measurement for $\Sigmac^{0,++}(2455)$ in hadronic collisions. The baryon-to-meson ratios $\Lambdac/\Dzero$ and $\Sigmac^{0,+,++}(2455)$/$\Dzero$ as well as the fraction of $\Lambdac$ feed-down from $\SigmacZeroPlusPlusPlus$ decays ($\LambdacFromScZeroPlusPlusPlus/\Lambdac$) are compared with expectations from the theoretical models described above. These ratios are calculated assuming the three $\Sigmac^{0,+,++}(2455)$ isospin states to be equally produced. In what follows, the symbols $\SigmacZeroPlusPlus$ and $\SigmacZeroPlusPlusPlus$ always refer to the ground-state $\SigmacZeroPlusPlusPlus(2455)$ baryons. 

The ALICE apparatus is described in detail in Refs.~\cite{Aamodt:2008zz, Abelev:2014ffa}. The $\Dzero$, $\Lambdac$, and $\SigmacZeroPlusPlus$ decays are reconstructed in the central barrel, which covers the pseudorapidity interval $|\eta|<0.9$ and is embedded in a cylindrical solenoid providing a magnetic field of 0.5\,T parallel to the beam direction. Charged particles are tracked with the Inner Tracking System (ITS) and the Time Projection Chamber (TPC). The ITS detector consists of six cylindrical silicon layers surrounding the beam pipe. The measurement of the specific energy loss (\dEdx) in the TPC gas and of the time difference between the collision time and the particle arrival time at the Time-Of-Flight (TOF) detector are exploited for particle identification (PID)~\cite{Adam:2016ilk,Acharya:2017kfy}. 

The data were collected with a minimum bias (MB) trigger requiring coincident signals in the two scintillator arrays covering the intervals $2.8<\eta<5.1$ (V0A) and $-3.7<\eta<-1.7$ (V0C). Only events with a primary vertex reconstructed within $\pm 10$\,cm from the nominal interaction point along the beam line were analysed. Events with multiple primary vertices were rejected in order to remove collision pileup in the same bunch crossing. The remaining undetected pileup is negligible. The selected events correspond to an integrated luminosity of $\mathcal{L}_{\rm int} = 31.9 \pm 0.5$\,${\rm nb}^{-1}$~\cite{aliceLumi13TeVrun2}. 

The following hadronic decay channels are reconstructed to measure the production of the $\SigmacZeroPlusPlus$, $\Lambdac$, and $\Dzero$ particles and their anti-particles. The $\SigmacZeroPlusPlus$ baryons decay strongly to a $\Lambdac$ in the channel $\SigmacZeroPlusPlus\to\pi^{-,+}\Lambdac$ with a branching ratio (BR) of about $100\%$~\cite{Zyla:2020zbs}. The $\Lambdac$ baryons are reconstructed in two different final states: $\LambdacTopKpi$, which occurs via multiple resonant and non-resonant decay channels, with a total ${\rm BR}$ of $(6.28\pm 0.32)\%$ and $\LambdacTopKzeroS$, with a ${\rm BR}$ of $(1.59\pm 0.08)\%$, followed by $\rm K^{0}_{S} \rightarrow \rm \pi^+\rm \pi^-$ with a ${\rm BR}$ of $(69.20 \pm 0.05)\%$. The $\Dzero$ mesons are reconstructed in the $\DtoKpi$ decay channel, which has a ${\rm BR}$ of $(3.95\pm 0.03)\%$.

The measurements of the $\Dzero$ and $\Lambdac$ cross sections are based on an invariant-mass analysis of signal candidates selected for having the proper daughter-particle identities and a displaced decay topology. The analysis procedure, described only briefly here, closely follows that of previous measurements~\cite{Acharya:2019mgn, Acharya:2017kfy,Acharya:2020lrg}. The $\Dzero$ candidates are formed by combining pairs of tracks with opposite charge, each with $|\eta| < 0.8$, $\pt > 0.3$\,\GeVc, and selected according to the track-quality criteria described in Ref.~\cite{Acharya:2019mgn}, which are adopted also in the $\Lambdac$ and $\SigmacZeroPlusPlus$ analyses. Pions and kaons are identified by requiring the   $\mathrm{d}E/\mathrm{d}x$ and time-of-flight measured respectively with the TPC and TOF to be within three times the detector resolution from the expected values. The topological selections applied to reduce the combinatorial background are the same as those reported in Ref.~\cite{Acharya:2019mgn}. For the $\LambdacTopKpi$ decay channel, $\Lambdac$ candidates are formed by combining tracks identified as p, K, or $\pi$, using the Bayesian PID approach with the \enquote{maximum-probability criterion}~\cite{Adam:2016acv}. The reconstruction of the $\LambdacTopKzeroS$ decay is based on a machine-learning classification that makes use of the Boosted Decision Trees (BDT) algorithm~\cite{Hocker:2007ht}. For both decay channels, a complete description of the applied PID and topological selections can be found in Ref.~\cite{Acharya:2017kfy}. A fiducial-acceptance selection ${|y|<y_{\rm fid}(\pt)}$ is applied to the $\Dzero$ and $\Lambdac$ candidates, with $y_{\rm fid}$ smoothly increasing from about 0.6 at $\pt=1$\,\GeVc to the maximum value of 0.8 at $\pt=5$\,\GeVc. 

 For the $\SigmacZeroPlusPlus$ study, separate analyses are carried out with candidates obtained from the two $\Lambdac$ decay channels: averages are then taken of the resulting cross sections and particle cross section ratios. For the study of the $\LambdacFromScZeroPlusPlus$ feed-down, the analysis is performed as a function of $\Lambdac$ $\pt$, rather than $\SigmacZeroPlusPlus$ $\pt$. The  $\SigmacZeroPlusPlus$ candidates are built by pairing $\Lambdac$ candidates with invariant mass in the interval $2.26\lesssim M(\Lambdac)\lesssim 2.31$\,\GeVmass with charged particles with $|\eta|<0.9$ and $\pt>0.12$\,\GeVc. The decay tracks are further selected for having a distance from the primary vertex smaller than 650\,$\mu$m in the transverse plane ($d_{r\varphi}$) and 1.5\,mm along the beam axis. The signal-to-background ratio for the $\SigmacZeroPlusPlus$ reconstructed with $\LambdacTopKpi$ candidates is improved by requiring $|d_{r\varphi}-d_{r\varphi}^\text{expected}|/\sigma(d_{r\varphi})<2.5$ for $4<\pt<6$\,\GeVc~\cite{Acharya:2017jgo}, and $\cos\theta_{\rm point}>0.8$ for $2<\pt<6$\,\GeVc, where $\theta_{\rm point}$ is the angle between the $\Lambdac$ flight line and its reconstructed momentum vector. 

\begin{figure}[t!]
\centering
\begin{minipage}{.45\textwidth}
  \centering
  \includegraphics[width=.99\linewidth,trim=1.8cm 0.4cm 0 -1.5cm]{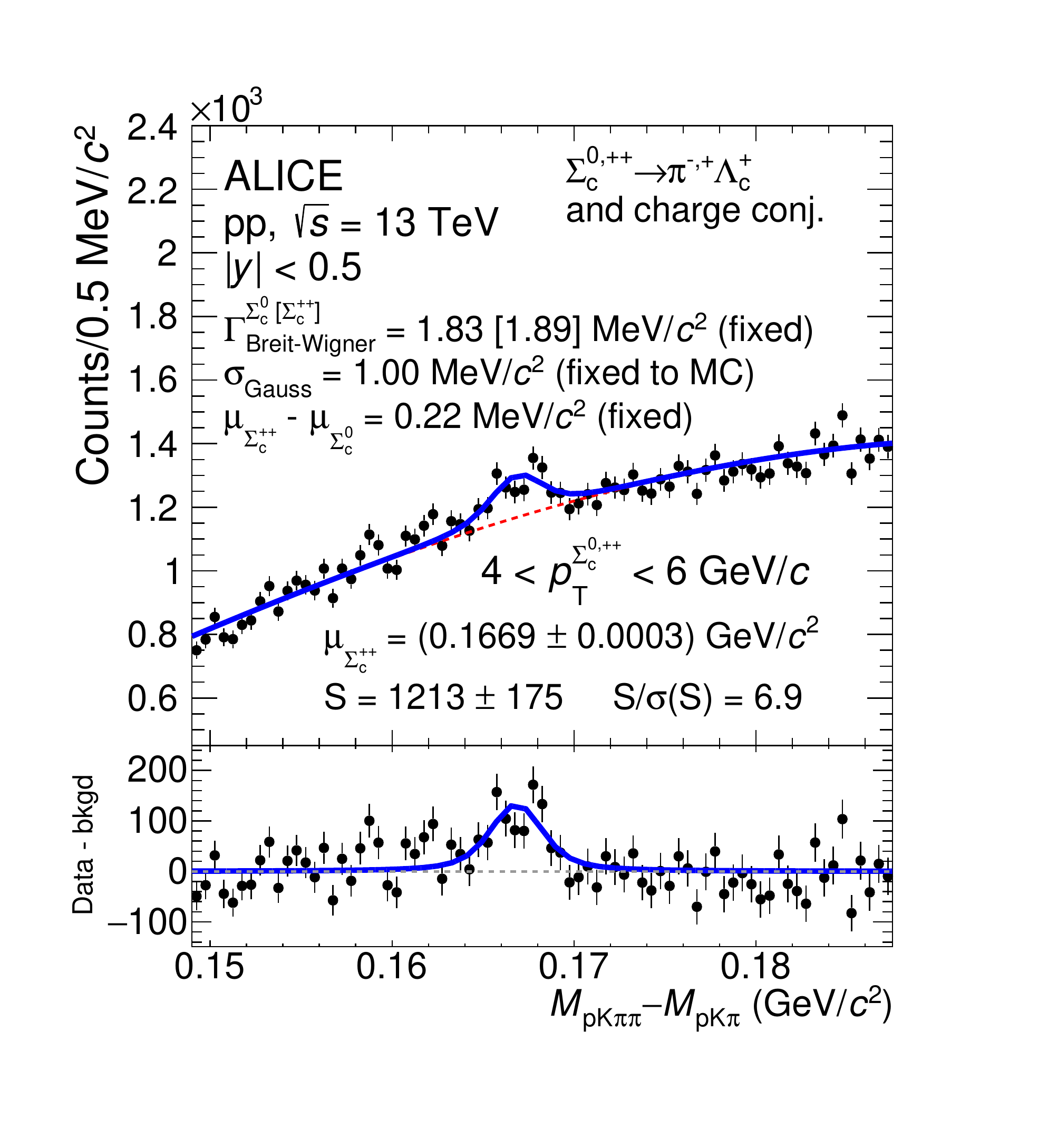} 
  \label{fig:test1}
\end{minipage}%
\begin{minipage}{.5\textwidth}
  \centering
  \includegraphics[width=0.99\linewidth,trim=2.5cm 0.4 0 -0cm]{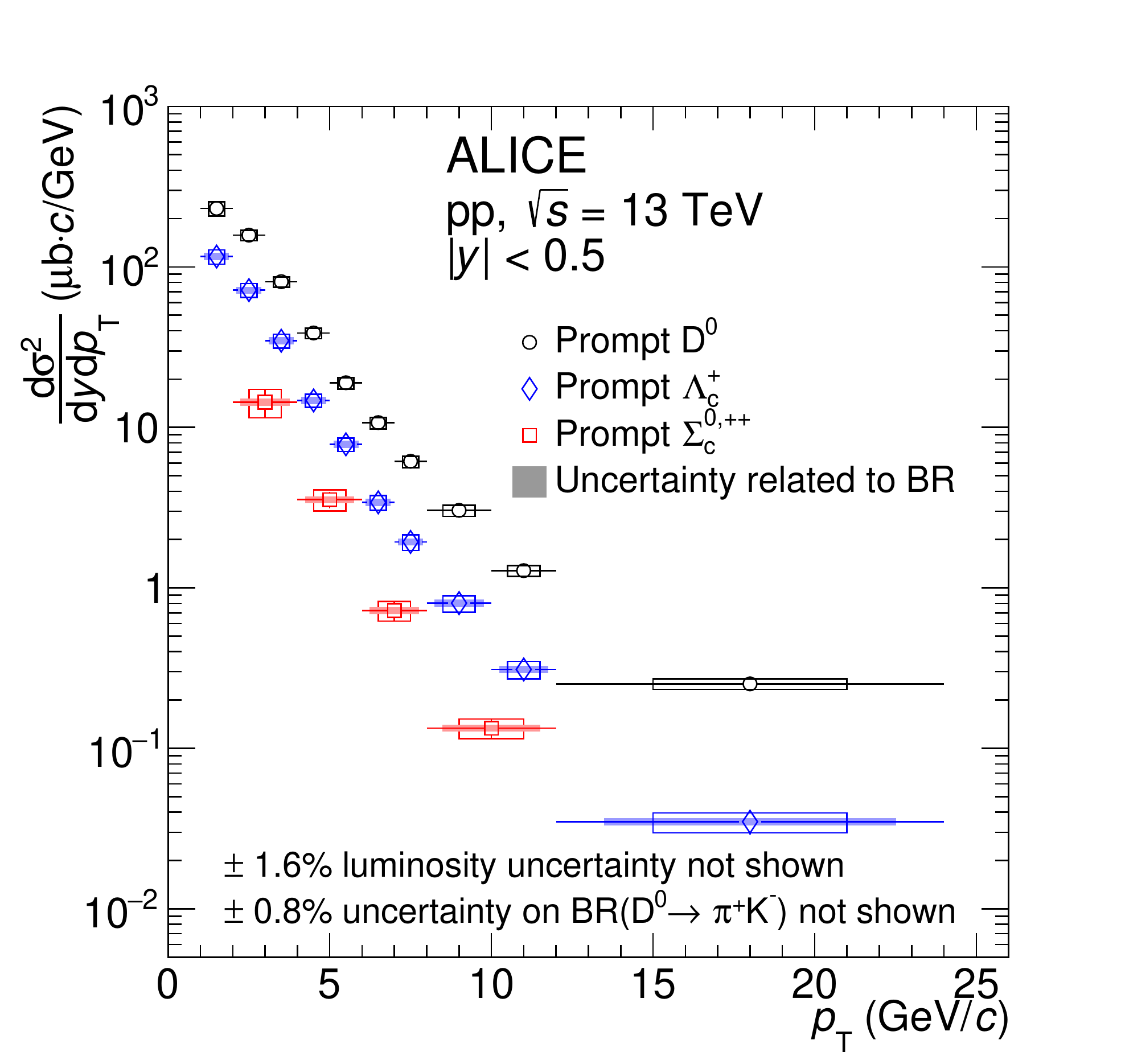} 
  \label{fig:test2}
\end{minipage}
\captionof{figure}{Left: distribution of $\pi^{+}\mathrm{K^{-}p}\pi^{\pm}$ to $\pi^{+}\mathrm{K^{-}p}$ (and charge conjugate) invariant-mass difference in $4<\ptSc<6$\,\GeVc. Right: $\pt$-differential cross section of prompt $\rm D^0$, $\Lambdac$, and $\SigmacZeroPlusPlus$ in pp collisions at $\sqrt{s}=13$\,\TeV. The statistical and systematic uncertainties are shown as vertical lines and boxes, respectively.}  
\label{fig:Massplot}
\end{figure}

The $\SigmacZeroPlusPlus$ and $\LambdacFromScZeroPlusPlus$ raw yields are estimated in each $\pt$ interval via a binned-likelihood fit to the distribution of the $\SigmacZeroPlusPlus$ and $\Lambdac$ candidate invariant-mass difference $\DeltaM$. An example of a $\DeltaM$ distribution is shown in Fig.~\ref{fig:Massplot} (left) for the $\LambdacTopKpi$ decay channel for $4<\ptScSigned<6$\,\GeVc. The function used to fit the signal peak is
\begin{equation}
    f(\Delta M)=\frac{C}{2}[\mathfrak{V}(\Delta M\!-\!\mu_{\SigmacPlusPlus};\sigma,\Gamma_{\SigmacPlusPlus})+\mathfrak{V}(\Delta M\!-\!\mu_{\SigmacPlusPlus}\!
    +
    \!\delta M;\sigma,\Gamma_{\SigmacZero})] \hspace{1mm},
\label{eq:signal_function}
\end{equation}
where $\mathfrak{V}$ is a Voigt function defined as the convolution of a Gaussian function and a Breit-Wigner function. Two Voigt functions are used to account for $\Sigma_{\mathrm c}^0$ ($M=2453.75\pm0.14$\,\MeVmass, full width $\Gamma_{\SigmacZero}=1.83^{+0.11}_{-0.19}$\,\MeVmass) and $\Sigma_{\rm c}^{++}$ ($M=2453.97\pm0.14$\,\MeVmass, $\Gamma_{\SigmacPlusPlus}=1.89^{+0.09}_{-0.18}$\,\MeVmass) isospin partners, whose invariant masses differ by $\delta M = 0.22$\,\MeVmass~\cite{Zyla:2020zbs}. The standard deviation of the Gaussian function, which accounts for the detector $\DeltaM$ resolution, is fixed to values $\sigma\sim 1$\,\MeVmass, determined from Monte Carlo simulations. The free parameters of the fit are $\mu_{\SigmacPlusPlus}$, i.e. the $\SigmacPlusPlus$ peak mean, and $C$, which represents the sum of $\SigmacZero$ and $\SigmacPlusPlus$ (and charge conjugates) raw yields. Depending on the $\pt$ interval, the background $\DeltaM$ distribution is described with a $3^{\rm rd}$-order polynomial function, a \enquote{threshold} function, or a template distribution, as described in Appendix~B. The statistical uncertainty of the raw yields varies between 15\% and 30\% depending on the decay channel and $\pt$ interval. It was verified that the $\SigmacZero$ and $\SigmacPlusPlus$ raw yields are compatible within statistical uncertainties. 

The $\pt$-differential cross sections of prompt $\Dzero$, $\Lambdac$, $\LambdacFromScZeroPlusPlus$, and $\SigmacZeroPlusPlus$ are calculated from the raw yields $N_{|y|<y_{\rm fid}}$, measured in the fiducial $y$ acceptance in a $\pt$ interval of width $\Delta \pt$, as
\begin{equation}
    \left.\frac{{\rm d}\sigma}{{\rm d}\pt}\right|_{|y|<0.5} =\frac{1}{2}\frac{1}{\Delta \pt}\!\times\frac{f_{\rm prompt}\times N_{|y|<y_{\rm fid}}}{c_{\Delta y}\times(A\times\varepsilon)_{\rm prompt}}\!\times\!\frac{1}{\rm BR}\!\times\!\frac{1}{\mathcal{L_{\rm int}}}\hspace{1mm}.
    \label{eq:cross_section} 
\end{equation}
The factor 2 in the denominator takes into account that both particles and antiparticles contribute to the measured raw yields. The term $c_{\Delta y}$ encompasses the correction for the rapidity coverage~\cite{Acharya:2017jgo}, and $(A\times\varepsilon)$ the detector acceptance as well as the reconstruction and selection efficiency for the signal. This is estimated from Monte Carlo simulations in which pp collisions are simulated with the PYTHIA 8.243~event generator~\cite{Sjostrand:2006za, Sjostrand:2014zea} and the generated particles are propagated through the apparatus using the GEANT3 package~\cite{Brun:1994aa} via a simulation that reproduces the detector layout and data-taking conditions. For prompt $\SigmacZeroPlusPlus$, $c_{\Delta y}\times(A\times\varepsilon)$ increases from 1\% (4\%) in $2<\pt<4$\,\GeVc to 11\% (22\%) in $8<\pt<12$\,\GeVc in the $\LambdacTopKpi$ ($\LambdacTopKzeroS$) analysis. 

The fraction of prompt particles contributing to the measured raw yield, $f_{\rm prompt}$, is calculated using the reconstruction efficiencies of prompt and feed-down signals and the feed-down $\Lambdac$ and $\Dzero$ cross sections, from $\Lambdab$ and B-meson decays (\enquote{beauty feed-down}). The latter cross sections are estimated as reported in Refs.~\cite{Acharya:2020lrg, Acharya:2019mgn}, using computations based on FONLL calculations~\cite{Cacciari:1998it,Cacciari:2012ny}, beauty-quark fragmentation fractions determined from LHCb data~\cite{Aaij:2019pqz} for $\mathrm{b}\rightarrow\Lambdab$ and from the averaged b-quark fragmentation fraction from LEP~\cite{Gladilin:2014tba} for ${\mathrm b\rightarrow \mathrm B}$, and modelling the $\Lambdab\to\Lambdac+{\it X}$ and $\rm{B}\to \Dzero + {\it X}$ decay kinematics with PYTHIA 8 simulations~\cite{Sjostrand:2007gs}. The values of $f_{\rm prompt}$ range from 0.8 to 0.96 depending on $\pt$ and the particle species. In the $\SigmacZeroPlusPlus$ case, according to currently known decays~\cite{Zyla:2020zbs} and to PYTHIA 8 simulations, a non-negligible feed-down contribution is only expected from $\Lambdab$ decays. The probability for $\Lambdab \to \SigmacZeroPlusPlus + {\it X}$ decays is estimated to be about 3\% of the probability for $\Lambdab\to\Lambdac+{\it X}$ decays, resulting in $f_{\rm prompt}\geq 95\%$ for both $\LambdacFromScZeroPlusPlus$ and $\SigmacZeroPlusPlus$ analyses. 

Several sources of systematic uncertainties of the measured cross sections were studied, following similar procedures to those described in Refs.~\cite{Acharya:2020lrg,Acharya:2019mgn} for the $\Lambdac$ and $\Dzero$ analyses. The uncertainty of $N_{|y|<y_{\rm fid}}$, estimated by varying the invariant mass fit procedure, ranges from 2\% to 4\% for $\Dzero$ and from 5\% to 11\% for $\Lambdac$, depending on $\pt$. For $\SigmacZeroPlusPlus$ and $\LambdacFromScZeroPlusPlus$, this source provides the largest contribution to the systematic uncertainty, which was estimated by repeating the $\DeltaM$ fits varying the signal and background fit functions, as well as the fit ranges. The $\Gamma$ and $\delta M$ parameters were varied within their uncertainties, and the Gaussian width $\sigma$ was changed by $\pm 20\%$. The estimated uncertainty decreases from 15--30\% in the first $\pt$ interval down to 8--10\% in the last one. Imperfections in the description of the apparatus and detector conditions in the Monte Carlo simulations introduce an uncertainty on the determination of the $c_{\Delta y}\times(A\times\varepsilon)_{\rm prompt}$ correction factor: the systematic uncertainty of the track-reconstruction efficiency induces an uncertainty of about 4\% for $\Dzero$, and 8\% for $\Lambdac$ and $\SigmacZeroPlusPlus$, while the uncertainty related to the signal-selection efficiency, estimated by varying both topological and PID selections, ranges between 3\% and 10\% depending on the $\pt$ interval and particle species. Variations of the simulated signal spectrum $\pt$ shapes based on FONLL (for $\Dzero$) and CR-BLC (for $\Lambdac$ and $\SigmacZeroPlusPlus$) models alter the efficiency by 2\% for $\Dzero$ with $\pt < 2$\,\GeVc and, for the other analyses, by values decreasing from 10\% to 1\% with increasing $\pt$. The systematic uncertainty of the prompt fraction is about 2--4\% for $\Dzero$ and $\Lambdac$. For the $\LambdacFromScZeroPlusPlusPlus$ and $\SigmacZeroPlusPlusPlus$ analyses, the beauty feed-down contribution was varied according to the $\Lambdac$ feed-down uncertainty, with the additional variation from 3\% to 6\% of the ratio of $\SigmacZeroPlusPlus$ and $\Lambdac$ feed-down estimated with PYTHIA 8 simulations as described previously. The resulting uncertainty of the cross section is within 2\%. Further $\pt$-independent uncertainties derive from the BR and the luminosity. All the uncertainty sources described above are assumed to be uncorrelated with respect to each other. The total uncertainty in each $\pt$ interval is calculated as the quadratic sum of the values estimated for each source. 
\begin{figure}
    \centering
     \includegraphics[width=1.0\textwidth]{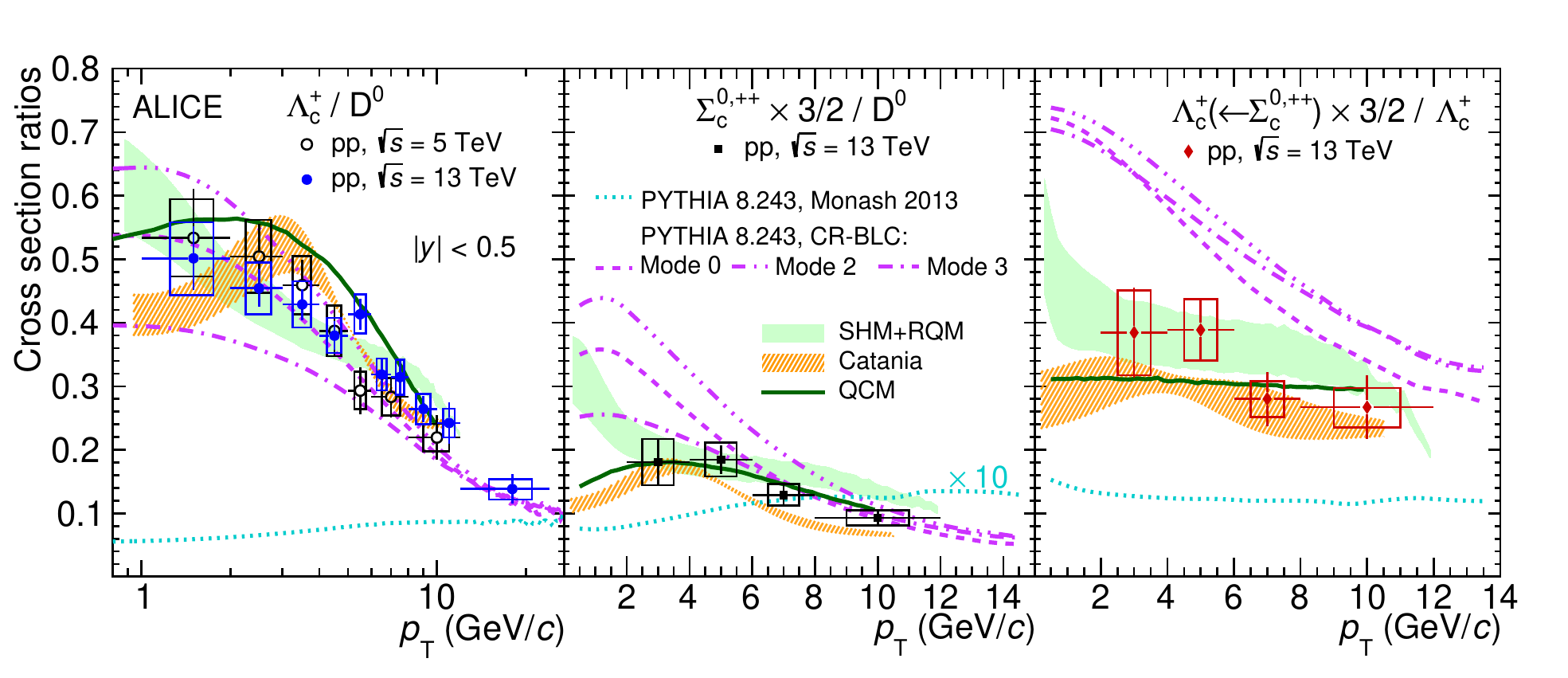}
    \caption{Prompt-charm-hadron cross-section ratios: $\Lambdac/\Dzero$ (left), $\SigmacZeroPlusPlusPlus/\Dzero$ (middle), and $\LambdacFromScZeroPlusPlusPlus/\Lambdac$ (right), in pp collisions at $\sqrt{s}=13$\,\TeV, compared with model expectations~\cite{Christiansen:2015yqa,He:2019tik,Plumari:2017ntm,Song:2018tpv} and (left) with data from pp collisions at $\sqrt{s}=5.02$\,\TeV~\cite{Acharya:2020uqi}. The horizontal lines reflect the width of the $\pt$ intervals. The PYTHIA Monash 2013 curve is scaled by a factor of 10 in the middle panel.}
    \label{fig:Ratios}
\end{figure}

The $\pt$-differential cross sections of $\Dzero$, $\Lambdac$, and $\SigmacZeroPlusPlus$ are shown in Fig.~\ref{fig:Massplot} (right). For $\Lambdac$ and $\SigmacZeroPlusPlus$ the weighted average of the results from the analyses of the two $\Lambdac$ decay channels is calculated, using the inverse of the quadratic sum of the relative statistical and uncorrelated systematic uncertainties as weights. The total systematic uncertainty of the averaged $\Sigmac$ cross section varies from 20\% at low \pt to 13\% at high \pt. The cross-section ratios $\Lambdac/\Dzero$ and $\SigmacZeroPlusPlusPlus/\Dzero$ are compared with model expectations in Fig.~\ref{fig:Ratios} (left and middle panels). In the ratios, the systematic uncertainties of the track-reconstruction efficiency and luminosity, considered as fully correlated, cancel partly and completely, respectively. The feed-down uncertainty is propagated as partially correlated, while all other uncertainties are treated as uncorrelated. The $\Lambdac/\Dzero$ ratio decreases with increasing $\pt$ and is significantly larger than the $\approx$0.12 values observed in $\eepm$ and $\ep$ collisions at several collision energies~\cite{Albrecht:1988an,Avery:1990bc,Albrecht:1991ss,Gladilin:2014tba,Chekanov:2005mm,Abramowicz:2013eja,Abramowicz:2010aa}. 
The values measured in pp collisions at $\sqrt{s}=13$\,\TeV are compatible, within uncertainties, with those measured at $\sqrts=5.02$\,\TeV~\cite{Acharya:2020lrg,Acharya:2020uqi}. As shown in Fig.~\ref{fig:Ratios} (middle), the $\SigmacZeroPlusPlusPlus/\Dzero$ ratio is close to 0.2 for $2<\pt<6$\,\GeVc, and decreases with $\pt$ down to about 0.1 for $8<\pt<12$\,\GeVc, though the uncertainties do not allow firm conclusions about the $\pt$ dependence to be made. From Belle measurements (Table~IV in Ref.~\cite{Niiyama:2017wpp}), the $\SigmacZeroPlusPlusPlus/\Lambdac$ ratio in $\eepm$ collisions at $\sqrt{s}=10.52$\,\GeV can be evaluated to be around 0.17 and, thus, the $\SigmacZeroPlusPlusPlus/\Dzero$ ratio can be estimated to be around 0.02. Therefore, a remarkable difference is present between the pp and $\eepm$ collision systems. Although rather approximate, such comparison is corroborated by the fact that a simulation performed with the default version of PYTHIA 6.2 reasonably reproduces Belle data~\cite{Niiyama:2017wpp}, while the default version of PYTHIA 8.243 (Monash 2013 tune) severely underpredicts ALICE data, despite the very similar modelling of charm fragmentation in the two simulations. Figure~\ref{fig:Ratios} (right) shows the ratio $\LambdacFromScZeroPlusPlusPlus/\Lambdac$ as a function of $\pt$, which quantifies the fraction of $\Lambdac$ feed-down from $\SigmacZeroPlusPlusPlus$. In order to better exploit the cancellation of correlated uncertainties, this is calculated as the weighted average of the ratios measured separately in the $\LambdacTopKpi$ and $\LambdacTopKzeroS$ decay channels. The $\pt$-integrated value in the measured $\pt>2$\,\GeVc interval is $0.38\pm 0.06(\text{stat})\pm 0.06(\text{syst})$, significantly larger than the ratio $\SigmacZeroPlusPlusPlus/\Lambdac\sim$ 0.17 from Belle data and the $\sim$0.13 expectation from PYTHIA 8 (Monash 2013) simulations. This indicates a larger increase for $\SigmacZeroPlusPlusPlus/\Dzero$ than for the direct-$\Lambdac/\Dzero$ ratio from $\eepm$ to pp collisions. The larger feed-down from $\SigmacZeroPlusPlusPlus$ partially explains the difference between the $\Lambdac/\Dzero$ ratios in pp and $\eepm$ collisions. 
 
As shown in Figure~\ref{fig:Ratios}, the CR-BLC (for which the three variations defined in Ref.~\cite{Christiansen:2015yqa} are considered), SHM+RQM, and Catania models describe, within uncertainties, both the $\Lambdac/\Dzero$ and $\SigmacZeroPlusPlusPlus/\Dzero$ ratios. The QCM model uses the $\Lambdac/\Dzero$ data in pp collisions at $\sqrt{s}=7$\,\TeV to set the total charm baryon-to-meson ratio, but it predicts correctly the $\LambdacFromScZeroPlusPlusPlus/\Lambdac$ and the $\pt$-shape of all ratios. The $\LambdacFromScZeroPlusPlusPlus/\Lambdac$ ratio does not show a $\pt$ trend as steep as that expected from the CR-BLC model, which significantly overestimates the $\Lambdac$ feed-down from $\SigmacZeroPlusPlusPlus$ at low $\pt$. Therefore, the data suggest that further tuning of the model parameters involving the reconnection of quarks via junction topologies is needed to possibly validate this as the mechanism reducing the assumed suppression of $\SigmacZeroPlusPlusPlus$ formation in $\eepm$ collisions~\cite{Christiansen:2015yqa,Niiyama:2017wpp}. In the Catania, QCM, and SHM+RQM models, no specific penalty factor affects the formation of $\Sigmac$ states. The fact that the SHM+RQM model reproduces both the $\Lambdac/\Dzero$ ratio and the fraction of $\Lambdac$ feed-down from $\SigmacZeroPlusPlusPlus$ may suggest that yet-unobserved higher-mass charm-baryon states exist and are formed more frequently in pp collisions than in $\eepm$ and $\ep$ collisions. Similarly, the success of the Catania and QCM models in reproducing the data may indicate that charm hadronisation in pp collisions involves coalescence of charm quark with light quarks.

The $\pt$-differential cross section of $\SigmacZeroPlusPlus$ has been measured in pp collisions at $\sqrt{s}=13$\,\TeV in the range $2<\pt <12$\,\GeVc, the first measurement in hadron--hadron collisions, together with the $\Lambdac$ and $\Dzero$ cross sections in the range $1<\pt <24$\,\GeVc. The charm baryon-to-meson cross section ratios were found to be larger than expectations based on $\eepm$ measurements. The reported results confirm previous observations at $\sqrt{s}=5.02$\,\TeV and $\sqrt{s}=7$\,\TeV for the $\Lambdac$ and show for the first time that the effect also extends to the $\SigmacZeroPlusPlus$. The feed-down from $\SigmacZeroPlusPlusPlus$ decays to $\Lambdac$ production amounts to $0.38\pm 0.06(\text{stat})\pm 0.06(\text{syst})$
in the range $2<\pt <12$\,GeV/$c$, which is significantly larger than measurements in $\eepm$ collisions. The results presented provide important constraints on models aiming at explaining the observed increase of charm baryons in a parton-rich environment, either increasing baryon-formation probability via enhanced colour reconnection or coalescence mechanisms, or assuming feed-down from yet-unobserved higher-mass baryon states.

%% file: fa_2021-05-12.tex

The ALICE Collaboration would like to thank all its engineers and technicians for their invaluable contributions to the construction of the experiment and the CERN accelerator teams for the outstanding performance of the LHC complex.
The ALICE Collaboration gratefully acknowledges the resources and support provided by all Grid centres and the Worldwide LHC Computing Grid (WLCG) collaboration.
The ALICE Collaboration acknowledges the following funding agencies for their support in building and running the ALICE detector:
A. I. Alikhanyan National Science Laboratory (Yerevan Physics Institute) Foundation (ANSL), State Committee of Science and World Federation of Scientists (WFS), Armenia;
Austrian Academy of Sciences, Austrian Science Fund (FWF): [M 2467-N36] and Nationalstiftung f\"{u}r Forschung, Technologie und Entwicklung, Austria;
Ministry of Communications and High Technologies, National Nuclear Research Center, Azerbaijan;
Conselho Nacional de Desenvolvimento Cient\'{\i}fico e Tecnol\'{o}gico (CNPq), Financiadora de Estudos e Projetos (Finep), Funda\c{c}\~{a}o de Amparo \`{a} Pesquisa do Estado de S\~{a}o Paulo (FAPESP) and Universidade Federal do Rio Grande do Sul (UFRGS), Brazil;
Ministry of Education of China (MOEC) , Ministry of Science \& Technology of China (MSTC) and National Natural Science Foundation of China (NSFC), China;
Ministry of Science and Education and Croatian Science Foundation, Croatia;
Centro de Aplicaciones Tecnol\'{o}gicas y Desarrollo Nuclear (CEADEN), Cubaenerg\'{\i}a, Cuba;
Ministry of Education, Youth and Sports of the Czech Republic, Czech Republic;
The Danish Council for Independent Research | Natural Sciences, the VILLUM FONDEN and Danish National Research Foundation (DNRF), Denmark;
Helsinki Institute of Physics (HIP), Finland;
Commissariat \`{a} l'Energie Atomique (CEA) and Institut National de Physique Nucl\'{e}aire et de Physique des Particules (IN2P3) and Centre National de la Recherche Scientifique (CNRS), France;
Bundesministerium f\"{u}r Bildung und Forschung (BMBF) and GSI Helmholtzzentrum f\"{u}r Schwerionenforschung GmbH, Germany;
General Secretariat for Research and Technology, Ministry of Education, Research and Religions, Greece;
National Research, Development and Innovation Office, Hungary;
Department of Atomic Energy Government of India (DAE), Department of Science and Technology, Government of India (DST), University Grants Commission, Government of India (UGC) and Council of Scientific and Industrial Research (CSIR), India;
Indonesian Institute of Science, Indonesia;
Istituto Nazionale di Fisica Nucleare (INFN), Italy;
Institute for Innovative Science and Technology , Nagasaki Institute of Applied Science (IIST), Japanese Ministry of Education, Culture, Sports, Science and Technology (MEXT) and Japan Society for the Promotion of Science (JSPS) KAKENHI, Japan;
Consejo Nacional de Ciencia (CONACYT) y Tecnolog\'{i}a, through Fondo de Cooperaci\'{o}n Internacional en Ciencia y Tecnolog\'{i}a (FONCICYT) and Direcci\'{o}n General de Asuntos del Personal Academico (DGAPA), Mexico;
Nederlandse Organisatie voor Wetenschappelijk Onderzoek (NWO), Netherlands;
The Research Council of Norway, Norway;
Commission on Science and Technology for Sustainable Development in the South (COMSATS), Pakistan;
Pontificia Universidad Cat\'{o}lica del Per\'{u}, Peru;
Ministry of Education and Science, National Science Centre and WUT ID-UB, Poland;
Korea Institute of Science and Technology Information and National Research Foundation of Korea (NRF), Republic of Korea;
Ministry of Education and Scientific Research, Institute of Atomic Physics and Ministry of Research and Innovation and Institute of Atomic Physics, Romania;
Joint Institute for Nuclear Research (JINR), Ministry of Education and Science of the Russian Federation, National Research Centre Kurchatov Institute, Russian Science Foundation and Russian Foundation for Basic Research, Russia;
Ministry of Education, Science, Research and Sport of the Slovak Republic, Slovakia;
National Research Foundation of South Africa, South Africa;
Swedish Research Council (VR) and Knut \& Alice Wallenberg Foundation (KAW), Sweden;
European Organization for Nuclear Research, Switzerland;
Suranaree University of Technology (SUT), National Science and Technology Development Agency (NSDTA) and Office of the Higher Education Commission under NRU project of Thailand, Thailand;
Turkish Energy, Nuclear and Mineral Research Agency (TENMAK), Turkey;
National Academy of  Sciences of Ukraine, Ukraine;
Science and Technology Facilities Council (STFC), United Kingdom;
National Science Foundation of the United States of America (NSF) and United States Department of Energy, Office of Nuclear Physics (DOE NP), United States of America.

%% file: 2021-05-12-Alice_Authorlist_2021-05-12.tex
%

\small
\begin{flushleft}

S.~Acharya$^{\rm 143}$, 
D.~Adamov\'{a}$^{\rm 98}$, 
A.~Adler$^{\rm 76}$, 
J.~Adolfsson$^{\rm 83}$, 
G.~Aglieri Rinella$^{\rm 35}$, 
M.~Agnello$^{\rm 31}$, 
N.~Agrawal$^{\rm 55}$, 
Z.~Ahammed$^{\rm 143}$, 
S.~Ahmad$^{\rm 16}$, 
S.U.~Ahn$^{\rm 78}$
% ...

\section*{Affiliation notes}

$^{\rm I}$ Deceased\\
$^{\rm II}$ Also at: Italian National Agency for New Technologies, Energy and Sustainable Economic Development (ENEA), Bologna, Italy\\
$^{\rm III}$ Also at: Dipartimento DET del Politecnico di Torino, Turin, Italy\\
$^{\rm IV}$ Also at: M.V. Lomonosov Moscow State University, D.V. Skobeltsyn Institute of Nuclear, Physics, Moscow, Russia\\
$^{\rm V}$ Also at: Institute of Theoretical Physics, University of Wroclaw, Poland\\

\section*{Collaboration Institutes}

$^{1}$ A.I. Alikhanyan National Science Laboratory (Yerevan Physics Institute) Foundation, Yerevan, Armenia\\
$^{2}$ AGH University of Science and Technology, Cracow, Poland\\
$^{3}$ Bogolyubov Institute for Theoretical Physics, National Academy of Sciences of Ukraine, Kiev, Ukraine\\
$^{4}$ Bose Institute, Department of Physics  and Centre for Astroparticle Physics and Space Science (CAPSS), Kolkata, India\\
$^{5}$ Budker Institute for Nuclear Physics, Novosibirsk, Russia\\
$^{6}$ California Polytechnic State University, San Luis Obispo, California, United States\\
$^{7}$ Central China Normal University, Wuhan, China\\
$^{8}$ Centro de Aplicaciones Tecnol\'{o}gicas y Desarrollo Nuclear (CEADEN), Havana, Cuba\\
$^{9}$ Centro de Investigaci\'{o}n y de Estudios Avanzados (CINVESTAV), Mexico City and M\'{e}rida, Mexico\\
$^{10}$ Chicago State University, Chicago, Illinois, United States\\
$^{11}$ China Institute of Atomic Energy, Beijing, China\\
$^{12}$ Chungbuk National University, Cheongju, Republic of Korea\\
$^{13}$ Comenius University Bratislava, Faculty of Mathematics, Physics and Informatics, Bratislava, Slovakia\\
$^{14}$ COMSATS University Islamabad, Islamabad, Pakistan\\
$^{15}$ Creighton University, Omaha, Nebraska, United States\\
$^{16}$ Department of Physics, Aligarh Muslim University, Aligarh, India\\
$^{17}$ Department of Physics, Pusan National University, Pusan, Republic of Korea\\
$^{18}$ Department of Physics, Sejong University, Seoul, Republic of Korea\\
$^{19}$ Department of Physics, University of California, Berkeley, California, United States\\
$^{20}$ Department of Physics, University of Oslo, Oslo, Norway\\
$^{21}$ Department of Physics and Technology, University of Bergen, Bergen, Norway\\
$^{22}$ Dipartimento di Fisica dell'Universit\`{a} 'La Sapienza' and Sezione INFN, Rome, Italy\\
$^{23}$ Dipartimento di Fisica dell'Universit\`{a} and Sezione INFN, Cagliari, Italy\\
$^{24}$ Dipartimento di Fisica dell'Universit\`{a} and Sezione INFN, Trieste, Italy\\
$^{25}$ Dipartimento di Fisica dell'Universit\`{a} and Sezione INFN, Turin, Italy\\
$^{26}$ Dipartimento di Fisica e Astronomia dell'Universit\`{a} and Sezione INFN, Bologna, Italy\\
$^{27}$ Dipartimento di Fisica e Astronomia dell'Universit\`{a} and Sezione INFN, Catania, Italy\\
$^{28}$ Dipartimento di Fisica e Astronomia dell'Universit\`{a} and Sezione INFN, Padova, Italy\\
$^{29}$ Dipartimento di Fisica e Nucleare e Teorica, Universit\`{a} di Pavia, Pavia, Italy\\
$^{30}$ Dipartimento di Fisica `E.R.~Caianiello' dell'Universit\`{a} and Gruppo Collegato INFN, Salerno, Italy\\
$^{31}$ Dipartimento DISAT del Politecnico and Sezione INFN, Turin, Italy\\
$^{32}$ Dipartimento di Scienze e Innovazione Tecnologica dell'Universit\`{a} del Piemonte Orientale and INFN Sezione di Torino, Alessandria, Italy\\
$^{33}$ Dipartimento di Scienze MIFT, Universit\`{a} di Messina, Messina, Italy\\
$^{34}$ Dipartimento Interateneo di Fisica `M.~Merlin' and Sezione INFN, Bari, Italy\\
$^{35}$ European Organization for Nuclear Research (CERN), Geneva, Switzerland\\
$^{36}$ Faculty of Electrical Engineering, Mechanical Engineering and Naval Architecture, University of Split, Split, Croatia\\
$^{37}$ Faculty of Engineering and Science, Western Norway University of Applied Sciences, Bergen, Norway\\
$^{38}$ Faculty of Nuclear Sciences and Physical Engineering, Czech Technical University in Prague, Prague, Czech Republic\\
$^{39}$ Faculty of Science, P.J.~\v{S}af\'{a}rik University, Ko\v{s}ice, Slovakia\\
$^{40}$ Frankfurt Institute for Advanced Studies, Johann Wolfgang Goethe-Universit\"{a}t Frankfurt, Frankfurt, Germany\\
$^{41}$ Fudan University, Shanghai, China\\
$^{42}$ Gangneung-Wonju National University, Gangneung, Republic of Korea\\
$^{43}$ Gauhati University, Department of Physics, Guwahati, India\\
$^{44}$ Helmholtz-Institut f\"{u}r Strahlen- und Kernphysik, Rheinische Friedrich-Wilhelms-Universit\"{a}t Bonn, Bonn, Germany\\
$^{45}$ Helsinki Institute of Physics (HIP), Helsinki, Finland\\
$^{46}$ High Energy Physics Group,  Universidad Aut\'{o}noma de Puebla, Puebla, Mexico\\
$^{47}$ Hiroshima University, Hiroshima, Japan\\
$^{48}$ Hochschule Worms, Zentrum  f\"{u}r Technologietransfer und Telekommunikation (ZTT), Worms, Germany\\
$^{49}$ Horia Hulubei National Institute of Physics and Nuclear Engineering, Bucharest, Romania\\
$^{50}$ Indian Institute of Technology Bombay (IIT), Mumbai, India\\
$^{51}$ Indian Institute of Technology Indore, Indore, India\\
$^{52}$ Indonesian Institute of Sciences, Jakarta, Indonesia\\
$^{53}$ INFN, Laboratori Nazionali di Frascati, Frascati, Italy\\
$^{54}$ INFN, Sezione di Bari, Bari, Italy\\
$^{55}$ INFN, Sezione di Bologna, Bologna, Italy\\
$^{56}$ INFN, Sezione di Cagliari, Cagliari, Italy\\
$^{57}$ INFN, Sezione di Catania, Catania, Italy\\
$^{58}$ INFN, Sezione di Padova, Padova, Italy\\
$^{59}$ INFN, Sezione di Pavia, Pavia, Italy\\
$^{60}$ INFN, Sezione di Roma, Rome, Italy\\
$^{61}$ INFN, Sezione di Torino, Turin, Italy\\
$^{62}$ INFN, Sezione di Trieste, Trieste, Italy\\
$^{63}$ Inha University, Incheon, Republic of Korea\\
$^{64}$ Institute for Gravitational and Subatomic Physics (GRASP), Utrecht University/Nikhef, Utrecht, Netherlands\\
$^{65}$ Institute for Nuclear Research, Academy of Sciences, Moscow, Russia\\
$^{66}$ Institute of Experimental Physics, Slovak Academy of Sciences, Ko\v{s}ice, Slovakia\\
$^{67}$ Institute of Physics, Homi Bhabha National Institute, Bhubaneswar, India\\
$^{68}$ Institute of Physics of the Czech Academy of Sciences, Prague, Czech Republic\\
$^{69}$ Institute of Space Science (ISS), Bucharest, Romania\\
$^{70}$ Institut f\"{u}r Kernphysik, Johann Wolfgang Goethe-Universit\"{a}t Frankfurt, Frankfurt, Germany\\
$^{71}$ Instituto de Ciencias Nucleares, Universidad Nacional Aut\'{o}noma de M\'{e}xico, Mexico City, Mexico\\
$^{72}$ Instituto de F\'{i}sica, Universidade Federal do Rio Grande do Sul (UFRGS), Porto Alegre, Brazil\\
$^{73}$ Instituto de F\'{\i}sica, Universidad Nacional Aut\'{o}noma de M\'{e}xico, Mexico City, Mexico\\
$^{74}$ iThemba LABS, National Research Foundation, Somerset West, South Africa\\
$^{75}$ Jeonbuk National University, Jeonju, Republic of Korea\\
$^{76}$ Johann-Wolfgang-Goethe Universit\"{a}t Frankfurt Institut f\"{u}r Informatik, Fachbereich Informatik und Mathematik, Frankfurt, Germany\\
$^{77}$ Joint Institute for Nuclear Research (JINR), Dubna, Russia\\
$^{78}$ Korea Institute of Science and Technology Information, Daejeon, Republic of Korea\\
$^{79}$ KTO Karatay University, Konya, Turkey\\
$^{80}$ Laboratoire de Physique des 2 Infinis, Ir\`{e}ne Joliot-Curie, Orsay, France\\
$^{81}$ Laboratoire de Physique Subatomique et de Cosmologie, Universit\'{e} Grenoble-Alpes, CNRS-IN2P3, Grenoble, France\\
$^{82}$ Lawrence Berkeley National Laboratory, Berkeley, California, United States\\
$^{83}$ Lund University Department of Physics, Division of Particle Physics, Lund, Sweden\\
$^{84}$ Moscow Institute for Physics and Technology, Moscow, Russia\\
$^{85}$ Nagasaki Institute of Applied Science, Nagasaki, Japan\\
$^{86}$ Nara Women{'}s University (NWU), Nara, Japan\\
$^{87}$ National and Kapodistrian University of Athens, School of Science, Department of Physics , Athens, Greece\\
$^{88}$ National Centre for Nuclear Research, Warsaw, Poland\\
$^{89}$ National Institute of Science Education and Research, Homi Bhabha National Institute, Jatni, India\\
$^{90}$ National Nuclear Research Center, Baku, Azerbaijan\\
$^{91}$ National Research Centre Kurchatov Institute, Moscow, Russia\\
$^{92}$ Niels Bohr Institute, University of Copenhagen, Copenhagen, Denmark\\
$^{93}$ Nikhef, National institute for subatomic physics, Amsterdam, Netherlands\\
$^{94}$ NRC Kurchatov Institute IHEP, Protvino, Russia\\
$^{95}$ NRC \guillemotleft Kurchatov\guillemotright  Institute - ITEP, Moscow, Russia\\
$^{96}$ NRNU Moscow Engineering Physics Institute, Moscow, Russia\\
$^{97}$ Nuclear Physics Group, STFC Daresbury Laboratory, Daresbury, United Kingdom\\
$^{98}$ Nuclear Physics Institute of the Czech Academy of Sciences, \v{R}e\v{z} u Prahy, Czech Republic\\
$^{99}$ Oak Ridge National Laboratory, Oak Ridge, Tennessee, United States\\
$^{100}$ Ohio State University, Columbus, Ohio, United States\\
$^{101}$ Petersburg Nuclear Physics Institute, Gatchina, Russia\\
$^{102}$ Physics department, Faculty of science, University of Zagreb, Zagreb, Croatia\\
$^{103}$ Physics Department, Panjab University, Chandigarh, India\\
$^{104}$ Physics Department, University of Jammu, Jammu, India\\
$^{105}$ Physics Department, University of Rajasthan, Jaipur, India\\
$^{106}$ Physikalisches Institut, Eberhard-Karls-Universit\"{a}t T\"{u}bingen, T\"{u}bingen, Germany\\
$^{107}$ Physikalisches Institut, Ruprecht-Karls-Universit\"{a}t Heidelberg, Heidelberg, Germany\\
$^{108}$ Physik Department, Technische Universit\"{a}t M\"{u}nchen, Munich, Germany\\
$^{109}$ Politecnico di Bari and Sezione INFN, Bari, Italy\\
$^{110}$ Research Division and ExtreMe Matter Institute EMMI, GSI Helmholtzzentrum f\"ur Schwerionenforschung GmbH, Darmstadt, Germany\\
$^{111}$ Russian Federal Nuclear Center (VNIIEF), Sarov, Russia\\
$^{112}$ Saha Institute of Nuclear Physics, Homi Bhabha National Institute, Kolkata, India\\
$^{113}$ School of Physics and Astronomy, University of Birmingham, Birmingham, United Kingdom\\
$^{114}$ Secci\'{o}n F\'{\i}sica, Departamento de Ciencias, Pontificia Universidad Cat\'{o}lica del Per\'{u}, Lima, Peru\\
$^{115}$ St. Petersburg State University, St. Petersburg, Russia\\
$^{116}$ Stefan Meyer Institut f\"{u}r Subatomare Physik (SMI), Vienna, Austria\\
$^{117}$ SUBATECH, IMT Atlantique, Universit\'{e} de Nantes, CNRS-IN2P3, Nantes, France\\
$^{118}$ Suranaree University of Technology, Nakhon Ratchasima, Thailand\\
$^{119}$ Technical University of Ko\v{s}ice, Ko\v{s}ice, Slovakia\\
$^{120}$ The Henryk Niewodniczanski Institute of Nuclear Physics, Polish Academy of Sciences, Cracow, Poland\\
$^{121}$ The University of Texas at Austin, Austin, Texas, United States\\
$^{122}$ Universidad Aut\'{o}noma de Sinaloa, Culiac\'{a}n, Mexico\\
$^{123}$ Universidade de S\~{a}o Paulo (USP), S\~{a}o Paulo, Brazil\\
$^{124}$ Universidade Estadual de Campinas (UNICAMP), Campinas, Brazil\\
$^{125}$ Universidade Federal do ABC, Santo Andre, Brazil\\
$^{126}$ University of Cape Town, Cape Town, South Africa\\
$^{127}$ University of Houston, Houston, Texas, United States\\
$^{128}$ University of Jyv\"{a}skyl\"{a}, Jyv\"{a}skyl\"{a}, Finland\\
$^{129}$ University of Kansas, Lawrence, Kansas, United States\\
$^{130}$ University of Liverpool, Liverpool, United Kingdom\\
$^{131}$ University of Science and Technology of China, Hefei, China\\
$^{132}$ University of South-Eastern Norway, Tonsberg, Norway\\
$^{133}$ University of Tennessee, Knoxville, Tennessee, United States\\
$^{134}$ University of the Witwatersrand, Johannesburg, South Africa\\
$^{135}$ University of Tokyo, Tokyo, Japan\\
$^{136}$ University of Tsukuba, Tsukuba, Japan\\
$^{137}$ Universit\'{e} Clermont Auvergne, CNRS/IN2P3, LPC, Clermont-Ferrand, France\\
$^{138}$ Universit\'{e} de Lyon, CNRS/IN2P3, Institut de Physique des 2 Infinis de Lyon , Lyon, France\\
$^{139}$ Universit\'{e} de Strasbourg, CNRS, IPHC UMR 7178, F-67000 Strasbourg, France, Strasbourg, France\\
$^{140}$ Universit\'{e} Paris-Saclay Centre d'Etudes de Saclay (CEA), IRFU, D\'{e}partment de Physique Nucl\'{e}aire (DPhN), Saclay, France\\
$^{141}$ Universit\`{a} degli Studi di Foggia, Foggia, Italy\\
$^{142}$ Universit\`{a} di Brescia, Brescia, Italy\\
$^{143}$ Variable Energy Cyclotron Centre, Homi Bhabha National Institute, Kolkata, India\\
$^{144}$ Warsaw University of Technology, Warsaw, Poland\\
$^{145}$ Wayne State University, Detroit, Michigan, United States\\
$^{146}$ Westf\"{a}lische Wilhelms-Universit\"{a}t M\"{u}nster, Institut f\"{u}r Kernphysik, M\"{u}nster, Germany\\
$^{147}$ Wigner Research Centre for Physics, Budapest, Hungary\\
$^{148}$ Yale University, New Haven, Connecticut, United States\\
$^{149}$ Yonsei University, Seoul, Republic of Korea\\

\bigskip 

\end{flushleft} 

%% file: SupplementalMat.tex
\section{Invariant-mass plots and fit procedures}

\begin{figure}[t!]
\centering
  \includegraphics[width=.48\linewidth,trim=1.8cm 0.4cm 0  -1.5cm]{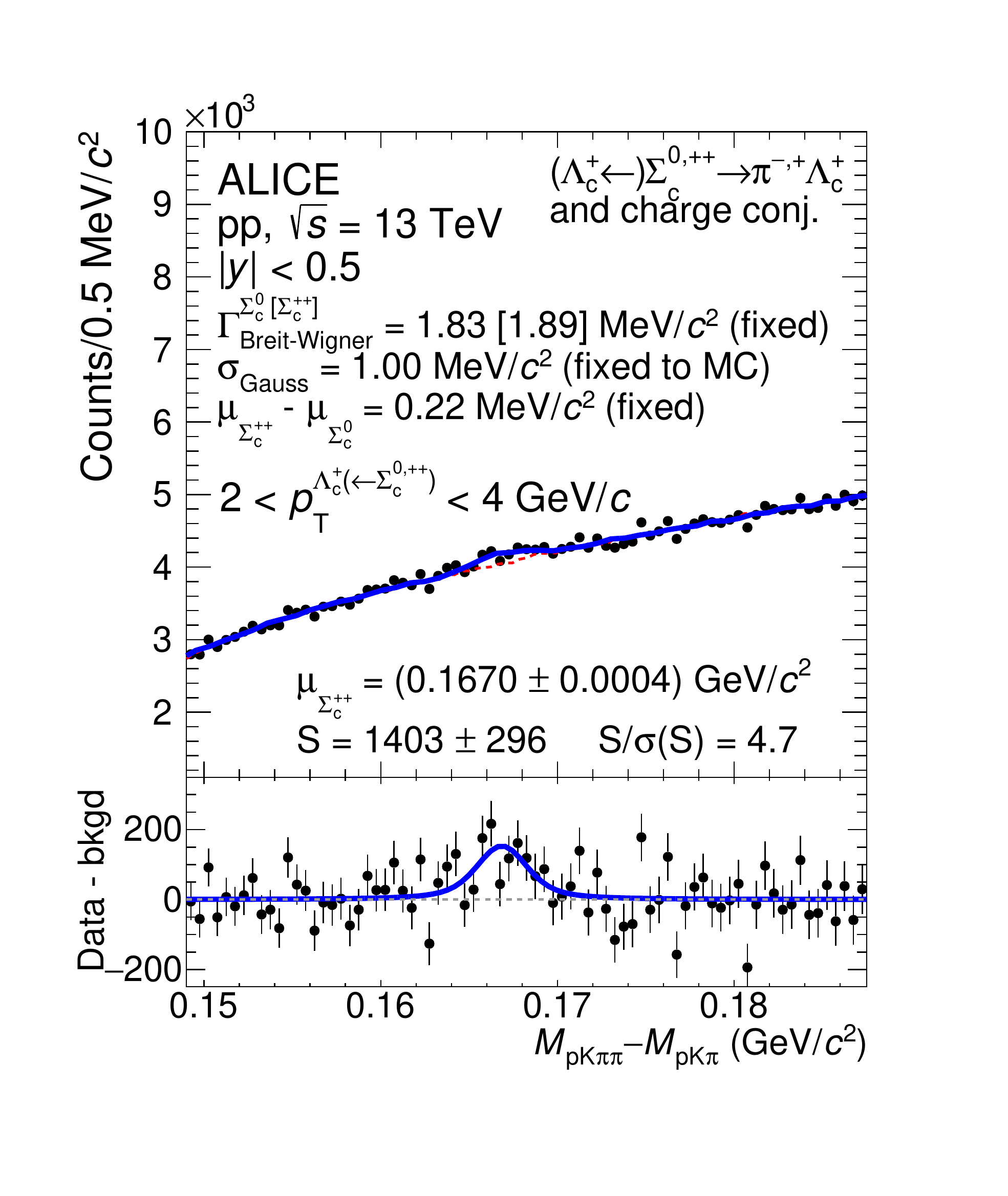}
  \includegraphics[width=.48\linewidth,trim=1.8cm 0.4cm 0 -1.5cm]{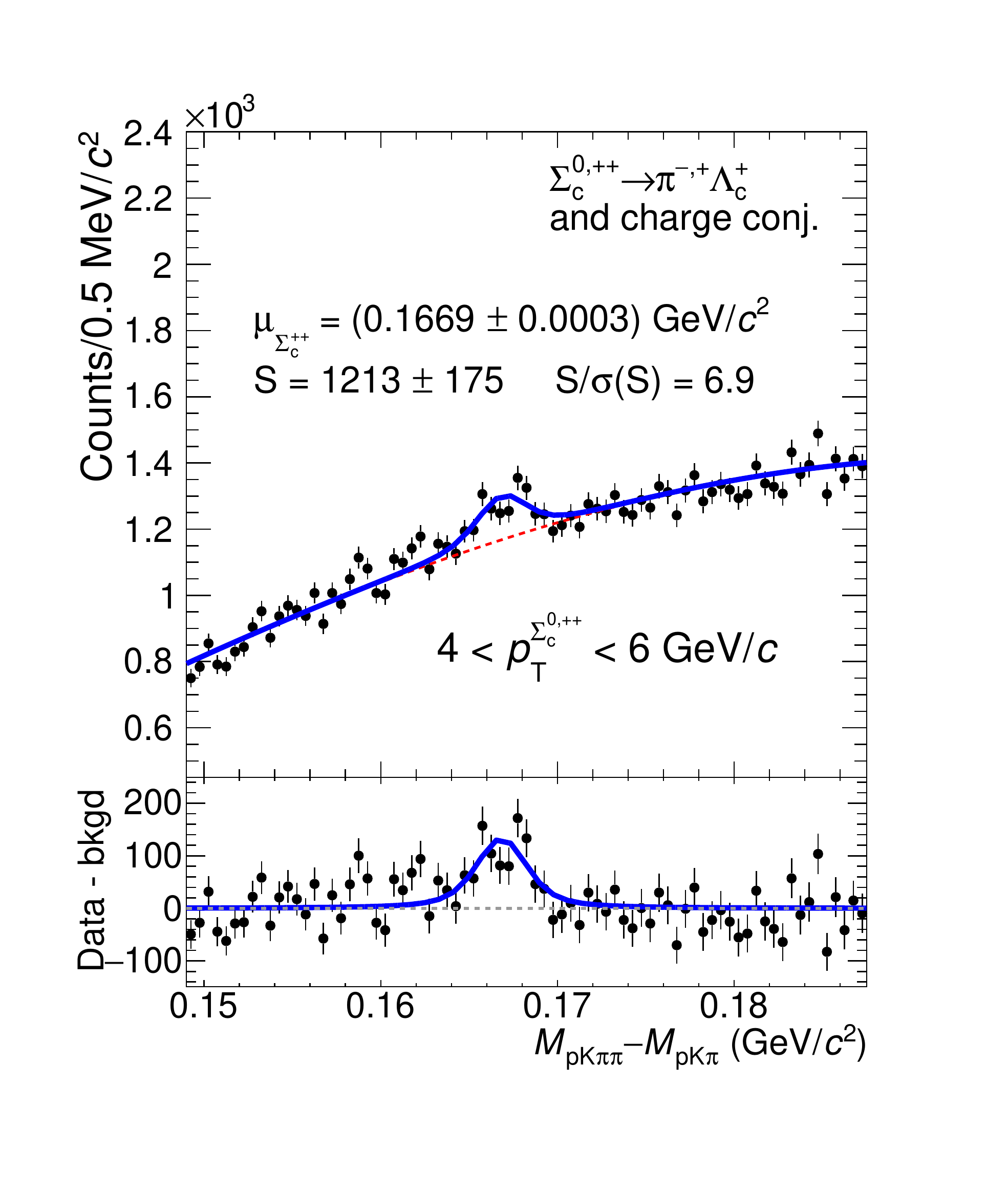} 
  \includegraphics[width=0.48\linewidth,trim=1.8cm 0.4 0 -0cm]{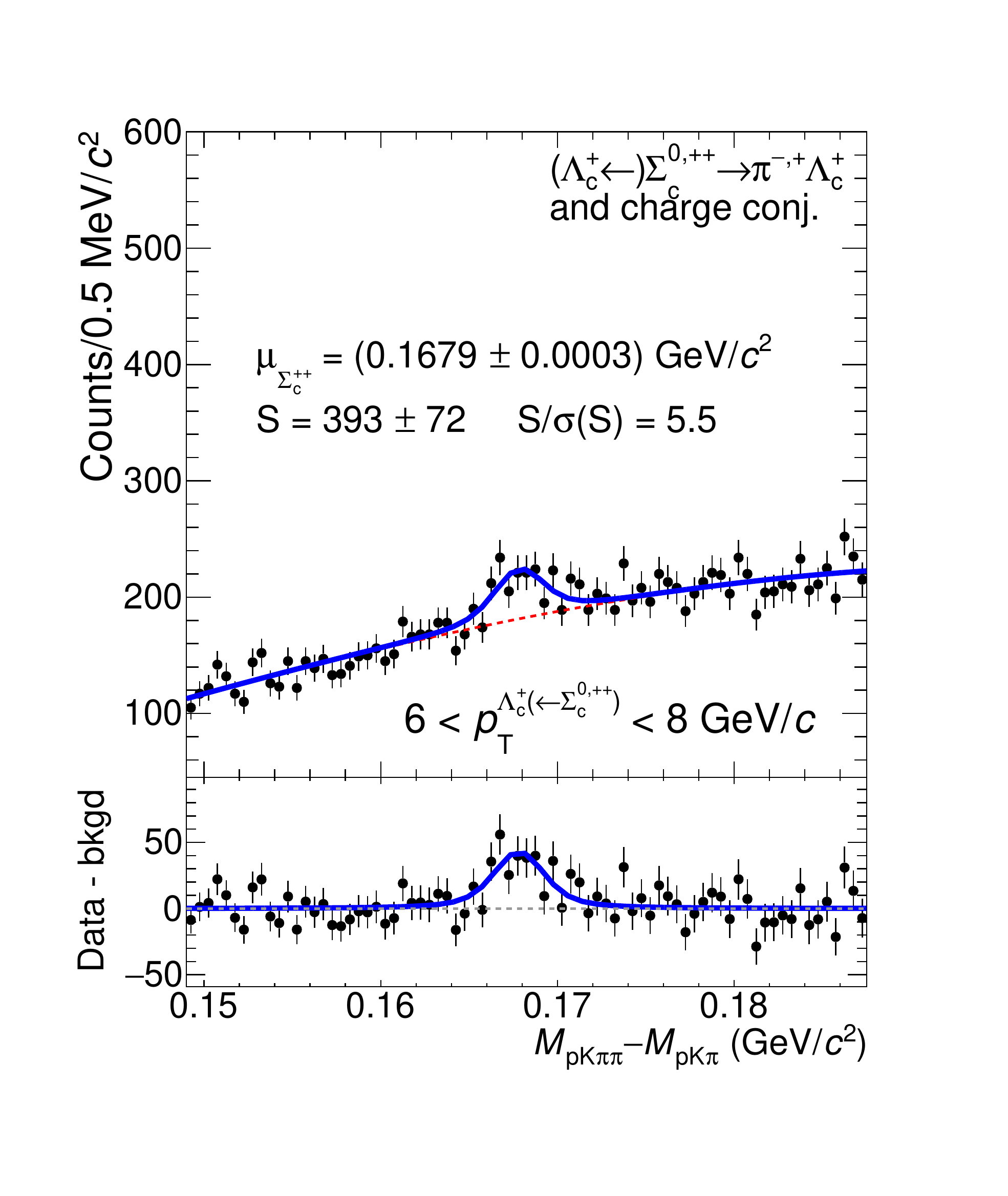}
    \includegraphics[width=.48\linewidth,trim=1.8cm 0.4cm 0 -1.5cm]{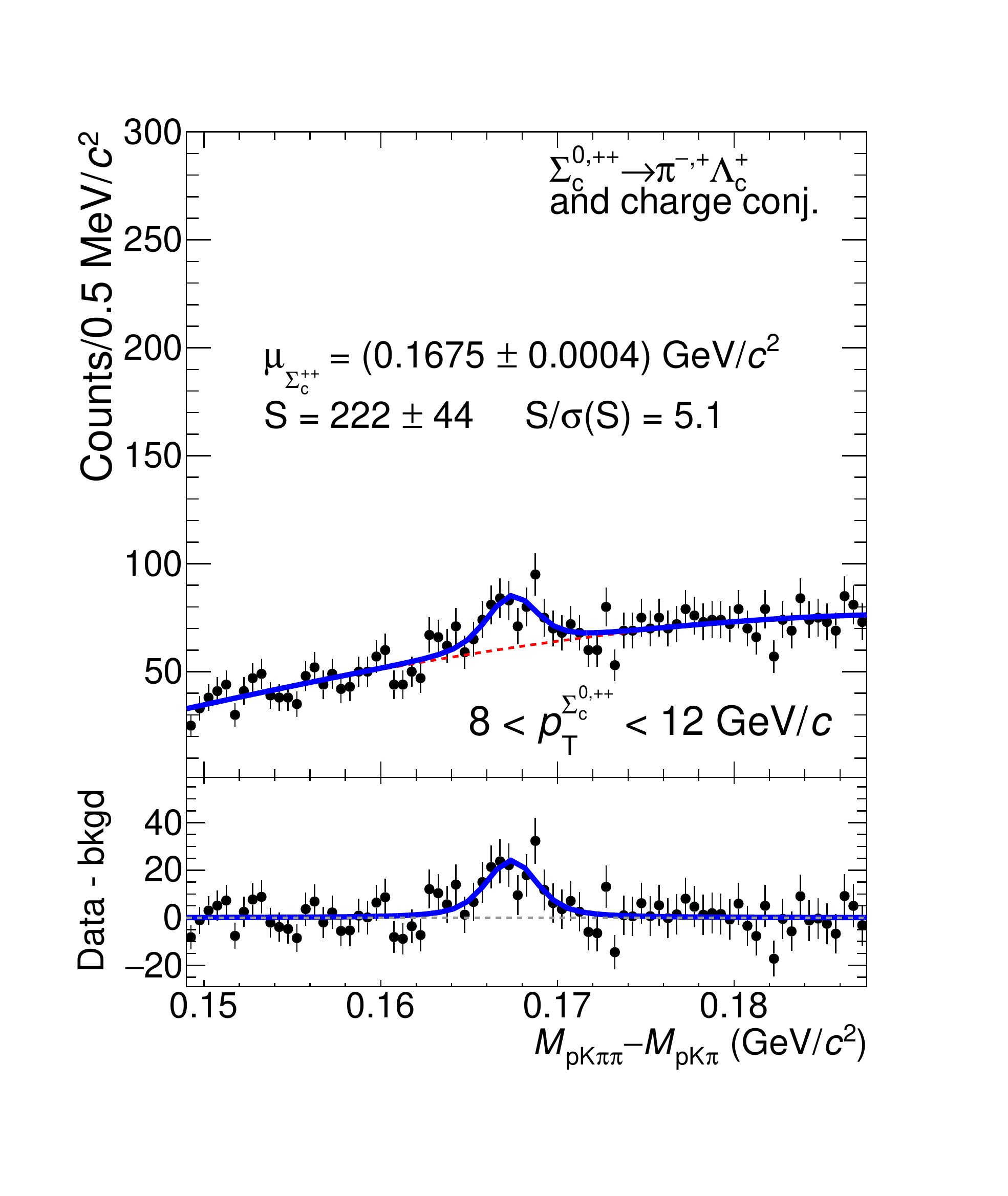}
\captionof{figure}{Distribution of $\mathrm{pK^{-}}\pi^+\pi^{\pm}$ to $\mathrm{pK^-}\pi^+$ (and charge conjugate) invariant-mass difference in different $\pt^{\LambdacFromScZeroPlusPlus}$ (left column) and $\ptSc$ (right column) intervals in pp collisions at $\sqrt{s}=13$\,\TeV. The residuals with respect to the background (\enquote{bkgd}) are shown in the bottom sub-panels.}  
\label{fig:MassplotSigmaCpkpi}
\end{figure}

\begin{figure}[t!]
\centering
  \includegraphics[width=.48\linewidth,trim=1.8cm 0.4cm 0  -1.5cm]{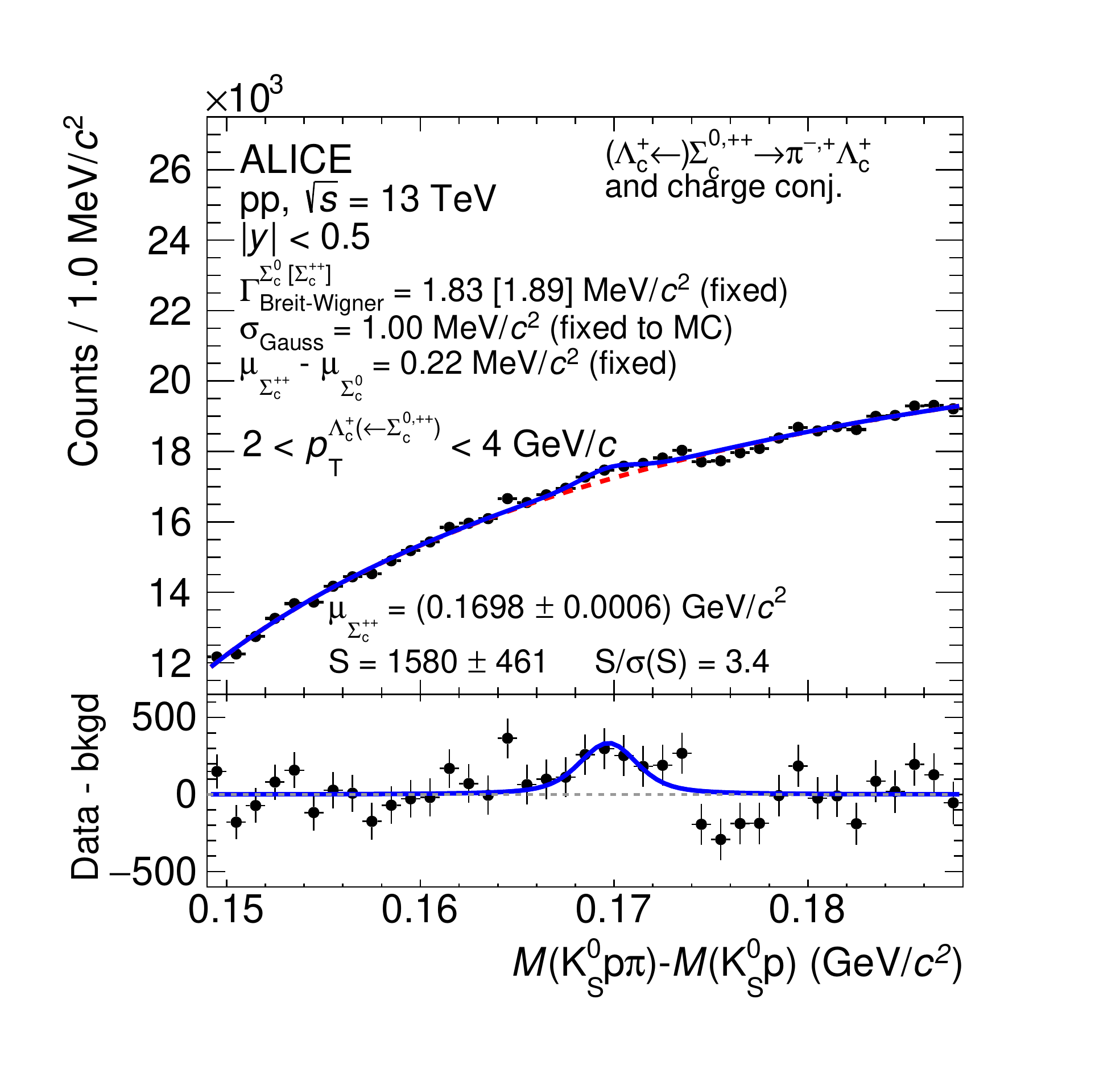}
  \includegraphics[width=.48\linewidth,trim=1.8cm 0.4cm 0 -1.5cm]{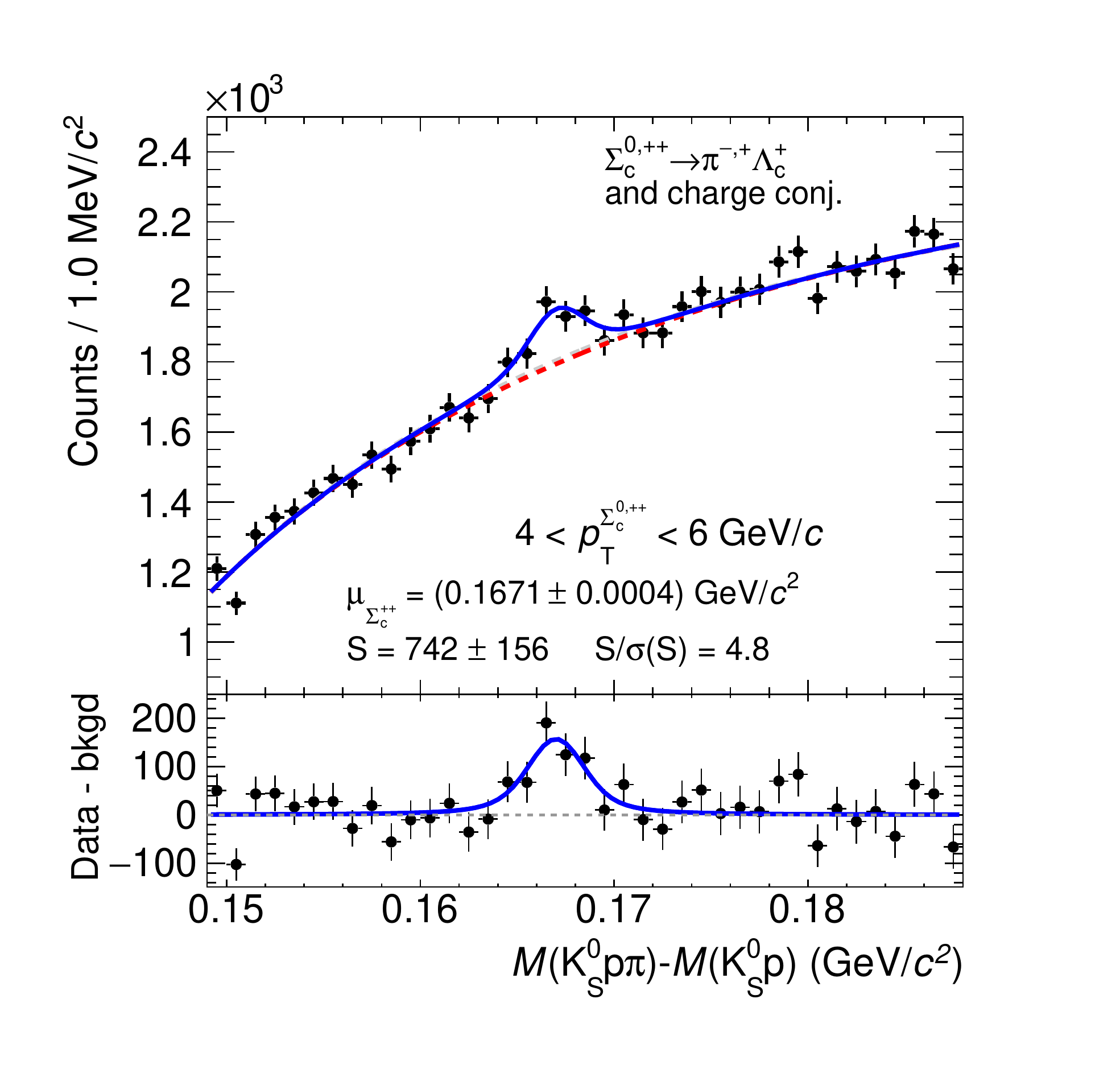}
  \includegraphics[width=0.48\linewidth,trim=1.8cm 0.4 0 -1.5cm]{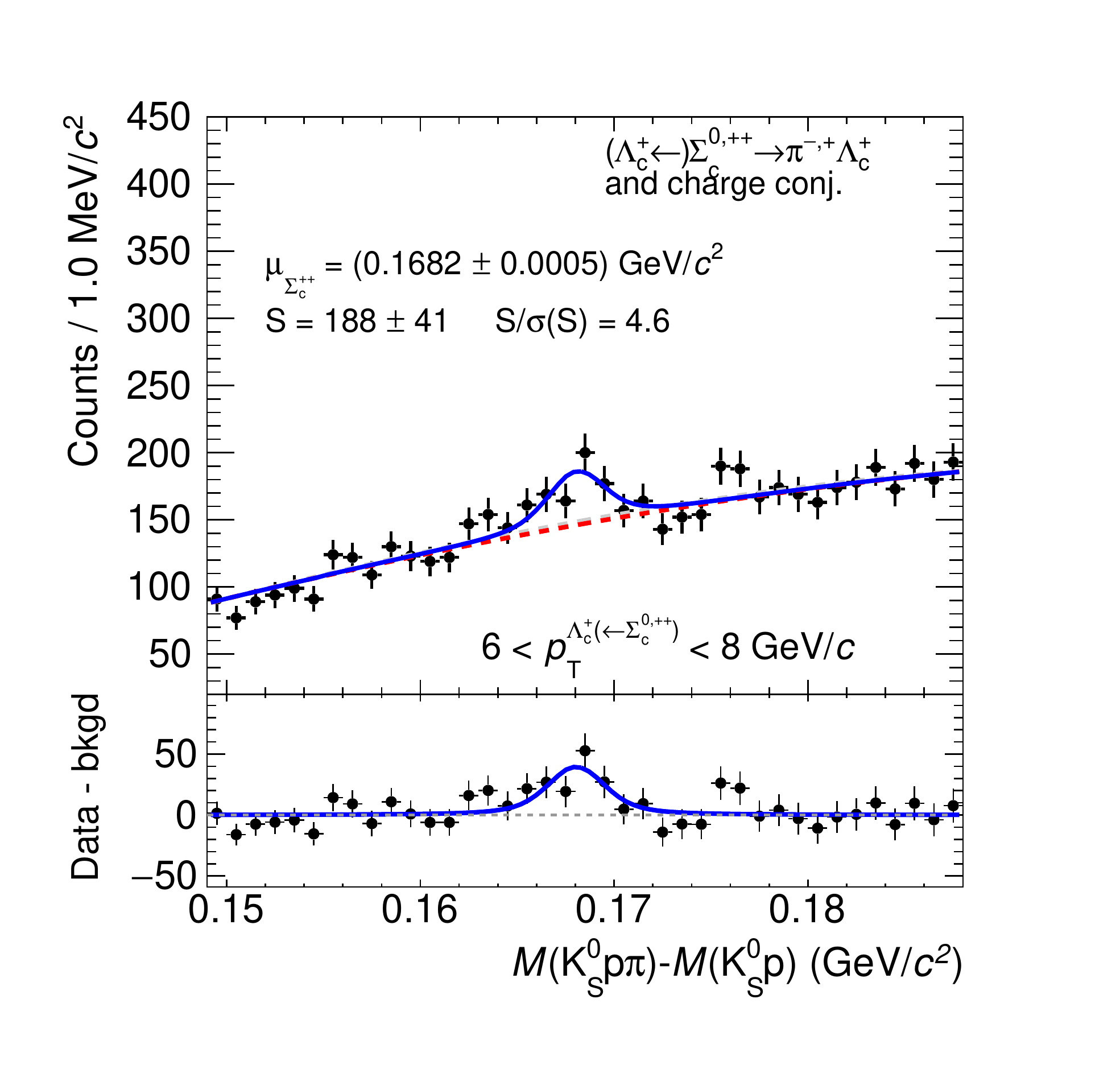}
  \includegraphics[width=.48\linewidth,trim=1.8cm 0.4cm 0 -1.5cm]{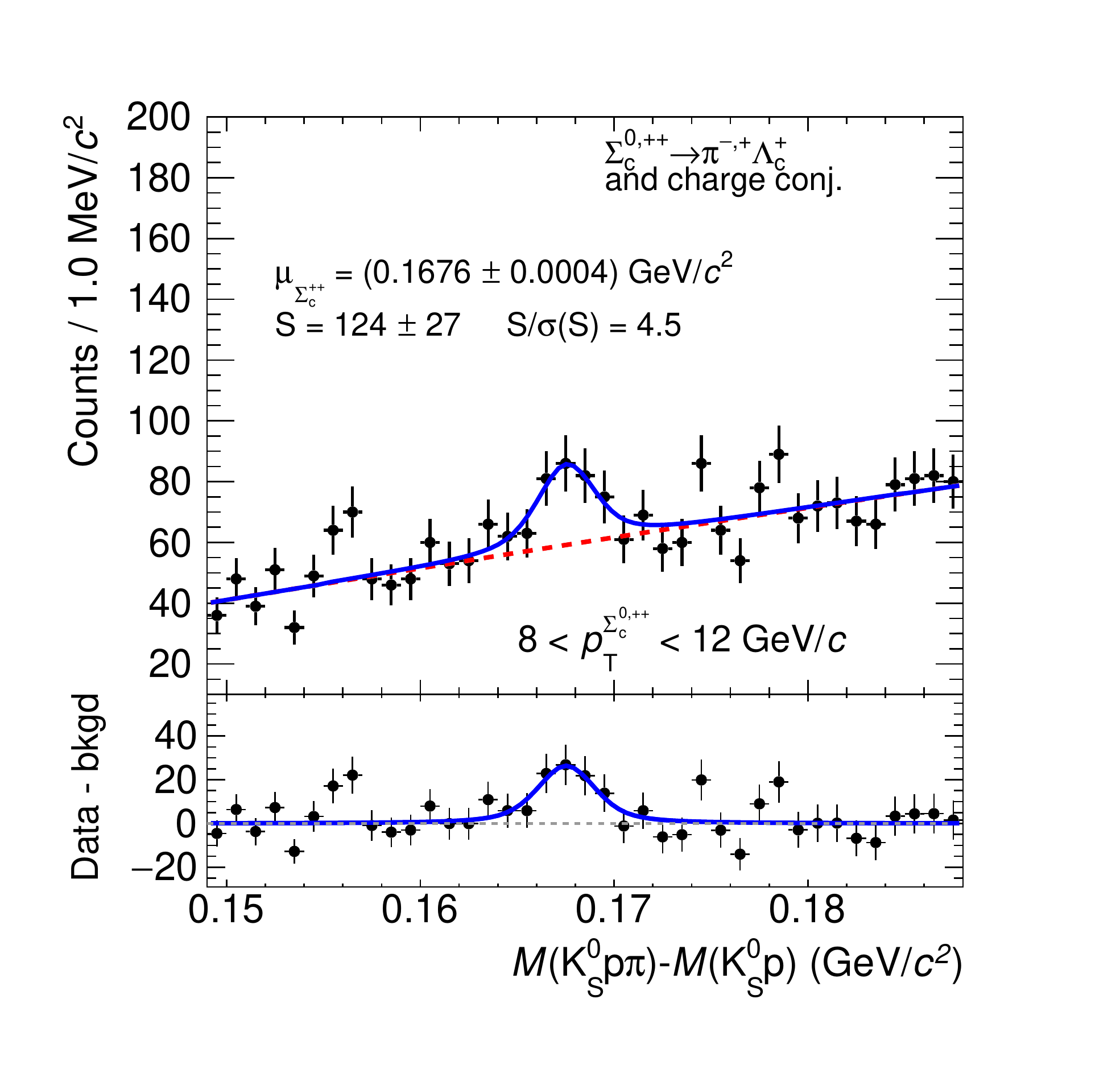}
\captionof{figure}{Distribution of $\mathrm{K^{0}_{S}p}\pi^\pm$ to $\mathrm{K^{0}_{S}p}$ (and charge conjugate) invariant-mass difference in different $\pt^{\LambdacFromScZeroPlusPlus}$ (left column) and $\ptSc$ (right column) intervals in pp collisions at $\sqrt{s}=13$\,\TeV. The residuals with respect to the background (\enquote{bkgd}) are shown in the bottom sub-panels.}  
\label{fig:MassplotSigmaCpK0s}
\end{figure}

The $\DeltaM$ distributions from which the $\SigmacZeroPlusPlus$ and $\LambdacFromScZeroPlusPlus$ raw yields are extracted and reported in Fig.~\ref{fig:MassplotSigmaCpkpi} for the analysis of the $\SigmacTopKzeropiS$ decay channel, and in Fig.~\ref{fig:MassplotSigmaCpK0s} for the analysis of the $\SigmacTopKpipi$ decay channel. 
Different functional shapes are used for the background term in the fit of the $\DeltaM$ distributions. 
For the analyses of the $\SigmacTopKzeropiS$ decay channel the \enquote{threshold} function ${\mathrm c_{0}}(\DeltaM - M_{\pi})^{\mathrm c_{1}}e^{-{\mathrm c_{2}}(\DeltaM-M_{\pi})}$ is used for $\pt<6$\,\GeVc, while a $3^{\rm rd}$-order polynomial function is used at higher $\pt$.
For the analyses of the $\SigmacTopKpipi$ decay channel, a $3^{\rm rd}$-order polynomial function is used for $\pt>4$\,\GeVc, while a template distribution multiplied by a $2^{\rm nd}$-order polynomial function is exploited in the interval $2<\pt<4$\,\GeVc, which is characterised by a very low signal-to-background ratio. The template distribution is obtained by recalculating $\DeltaM$ nine times for each $\SigmacZeroPlusPlus$ candidate after rotating the pion momentum vector around the longitudinal direction.
For both the $\SigmacTopKpipi$ and $\SigmacTopKzeropiS$ analyses the same background functional shapes are used in the fits of the $\SigmacZeroPlusPlus$ and $\LambdacFromScZeroPlusPlus$.

The difference obtained by using the template distribution for background candidates, the $3^{\rm rd}$-order polynomial and the \enquote{threshold} function was taken into account in the estimate of the systematic uncertainty on the raw-yield extraction.
\clearpage

\clearpage